\documentclass[12pt,thmsa,a4paper,notitlepage]{article}
\usepackage{amsfonts}

\input{tcilatex}

\begin{document}

\author{M. H. LAGRAA$^{1,2}$\thanks{%
e-mail : meriem.lagraa@gmail.com } and M. LAGRAA$^{2}$\thanks{%
e-mail : m.lagraa@lycos.com and lagraa.m@univ-oran.dz} \\
%EndAName
$^{1}$\'{E}cole sup\'{e}rieure en G\'{e}nie \'{E}lectrique et \'{E}nerg\'{e}%
tique d'Oran (ESG2E)\\
(Ex-EPSTO), \textit{B.P. 64 CH2 Achaba Hanifi, Oran, ALGERIA. }\\
$^{2}$\textit{Laboratoire de physique th\'{e}orique d'Oran (LPTO), }\\
\textit{Universit\'{e} d'Oran I, Ahmed Benbella, } \textit{B.P. 1524, El
M'Naouer,}\\
\textit{\ 31000 Es-S\'{e}nia, Oran, ALGERIA.}}
\title{On The Hamiltonian Formalism Of The Tetrad-Gravity With Fermions}
\maketitle

\begin{abstract}
We extend the analysis of the Hamiltonian formalism of the d-dimensional
tetrad-connection gravity to the fermionic field by fixing the non-dynamic
part of the spatial connection to zero \cite{Lagraa}. Although the reduced
phase space is equipped with complicated Dirac brackets, the first-class
constraints which generate the diffeomorphisms and the Lorentz
transformations satisfy a closed algebra with structural constants analogous
to that of the pure gravity. We also show the existence of a canonical
transformation leading to a new reduced phase space equipped with Dirac
brackets having a canonical form leading to the same algebra of the
first-class constraints.


\end{abstract}

\section{Introduction}

It is well-known that along the development of various canonical gravity
formalisms, the interest on incorporating matter to the theory has immensely
increased. In \cite{Nelson}, the analysis of the covariant Hamiltonian of
the tetrad-gravity coupled to the fermionic field was carried out in the
second order formalism. Note that in the presence of fermions the torsion is
not zero on-shell. This means that the first order formalism is not
equivalent to the correspondent second order Lagrangian obtained by
substituting the torsion-free spin connection in place of the connection,
unless modifying the action by an appropriate term \cite{Lagraa1}.

The first order Hamiltonian formalism of the Ashtekar complex connection 
\cite{Ashtekar} coupled to the fermionic matter has been done in \cite%
{Jacobson}. To avoid the complex nature of the Ashtekar connection the
action is modified by the Holst term \cite{holst} depending on a new
parameter known as the Barbero-Immirzi parameter \cite{Barb}. This permits
the construction of the Hilbert space in loop quantum gravity by reducing,
in the time gauge, the Lorentz's manifest invariance of the action to the
compact SO(3) subgroup \cite{Ashtekar1}. The Holst term does not affect the
classical equation of motion but this unphysical parameter appears in the
expressions of the observables at the quantum level \cite{Rovelli1}. Even at
the classical level the effect of this unphysical parameter must be observed
via the no vanishing torsion which emerges in the effective action under the
form of current-current interaction \cite{Perez}. In\ the time gauge of the $%
su(2)$ valued Ashtekar-Barbero connection formalism, a detailed canonical
treatment of non-minimal coupling of fermion generalizing the results of 
\cite{Mercury} is done in \cite{Bojowald}. All these works start from the
very beginning by the A.D.M. decomposition of the tetrad components in terms
of lapse and shift. This leads to an algebra of first-class constraints
involving structure functions meaning that the symmetries generated by these
constraints are not based on true Lie groups.

Even in the case where the A.D.M. phase space is extended to the Lagrange
multipliers (the lapse and the shift) as in \cite{Thiemann}, the analysis of
the covariant Hamiltonian formalism of super gravity theories leads to an
algebra of constraints which closes with structure functions.

In this paper we extent the Hamiltonian formalism of $d-$dimensional
tetrad-gravity minimally coupled to the fermionic field, where the
non-dynamic part of the spatial connection is fixed to zero \cite{Lagraa}.
This analysis is performed without the A.D.M.\ decomposition of the tetrad
components and is free of the Barbero-Immirzi parameter which is peculiar to
the $4-$dimension. In this framework we show the consistency of the
Hamiltonian formalism and establish the Dirac brackets of the reduced phase
space from which we derive the algebra of the first-class constraints which
closes with structure constants.

The paper is organized as follows: In section II we start with the action of
the tetrad-gravity coupled with the fermionic field where the non-dynamic
part of the spacial connection is fixed to zero. This hypothesis permits us
to get a consistent Hamiltonian formulation of the tetrad gravity coupled
with the fermionic field. In section III we establish the Dirac brackets of
the reduced phase space elements where the second-class constraints are
eliminated as strong equalities. We show that even if the Dirac brackets of
the connection with itself and with the fermionic field have complicated non
polynomial expressions, the reduced first-class constraints are polynomial
and satisfy the same closed algebra (with structural constants) as that of
the pure gravity.\ We show the existence of a canonical transformation
leading to a new reduced phase space which is canonical in terms of the
Dirac brackets. The first-class constraints defined on this new canonical
phase space satisfy the same closed algebra with structure constants. In the
appendix A we collect the properties of the projectors which allows us to
fix the non-dynamic projected part of the connection to zero.

\section{Hamiltonian formalism}

In this chapter we will extent the analysis of the Hamiltonian formalism of
the $d-$dimensional tetrad-gravity to the fermionic field by fixing the
non-dynamic part of the spatial connection to zero. The fixing of the
non-dynamic part of the connection to zero is necessary to avoid the
constraints resulting from the evolution equations of the non-dynamic part
of the spatial connection which are difficult to analyze \cite{Lagraa}.

As usual, to pass to the Hamiltonian formalism, we will suppose that the
manifold $\mathcal{M}$ has topology $R\times \Sigma $, where $t\in R$
represents the time which is the evolution parameter of $d-1$ dimensional
hypersurfaces $\Sigma _{t}$ in the $d-$dimensional manifold $\mathcal{M}$.\
We perform the Legendre transformations from the action $S(e,\omega ,\Psi )$ 
\begin{eqnarray}
S(e,\omega _{1},\Psi ) &=&\dint_{\mathcal{M}}\left( eA^{aKtL}\left( \partial
_{t}\omega _{1aKL}-D_{1a}\omega _{tKL}\right) -eA^{aKbL}\frac{\Omega _{1abKL}%
}{2}\right) d^{d}x  \nonumber \\
&&+\dint_{\mathcal{M}}ee^{tI}\frac{i}{2}\left( \overline{\Psi }\gamma
_{I}\partial _{t}\Psi -(\partial _{t}\overline{\Psi })\gamma _{I}\Psi
\right) d^{d}x  \nonumber \\
&&+\dint_{\mathcal{M}}\left( ee^{aI}\frac{i}{2}\left( \overline{\Psi }\gamma
_{I}D_{a}\Psi -(D_{a}\overline{\Psi })\gamma _{I}\Psi \right) -em\overline{%
\Psi }\Psi \right) d^{d}x  \nonumber \\
&&+\dint_{\mathcal{M}}ee^{tI}\frac{i}{2}\left( \overline{\Psi }(\gamma _{I}%
\frac{\sigma ^{KL}}{2}+\frac{\sigma ^{KL}}{2}\gamma _{I})\Psi \right) \omega
_{tKL}d^{d}x  \label{LagrangianF}
\end{eqnarray}%
where the spatial connection $\omega _{aKL}$ valued in $so(1,d-1)$
Lie-algebra is restricted to its dynamic part $\omega _{1aKL}=P_{1KaL}^{%
\text{ \ \ \ \ \ \ \ }PdQ}\omega _{dPQ}$ by fixing its non-dynamic part $%
\omega _{2aKL}=P_{2KaL}^{\text{ \ \ \ \ \ \ \ }PdQ}\omega _{dPQ}$ to zero
(A.10) where $P_{1KaL}^{\text{ \ \ \ \ \ \ \ }PdQ}$ and $P_{2KaL}^{\text{ \
\ \ \ \ \ \ }PdQ}$ are projectors (A.9). The covariant derivative is
expressed in terms of the dynamic part of the connection, $D_{1a}\omega
_{tKL}=P_{1KaL}^{\text{ \ \ \ \ \ \ \ }PdQ}D_{d}\omega _{tPQ}$ and $\Omega
_{1abKL}=\partial _{a}\omega _{1bKL}-\partial _{b}\omega _{1aKL}+\omega
_{1aK}^{\text{ \ \ \ \ \ }N}\omega _{1bNL}-\omega _{1bK}^{\text{ \ \ \ \ \ }%
N}\omega _{1aNL}$ is the curvature of $\omega _{1aKL}$.\ $x^{\mu }$ are
local coordinates of the $d-$dimensional manifold $\mathcal{M}$, the Greek
letters $\mu ,$ $\nu $ $\in \left[ 0,1,..,d-1\right] $ denote space-time
indices ($t$ represents the time $t=x^{t}=x^{0}$ and $x^{a}$ with the space
indices $a,b\in \left[ 1,..,d-1\right] $ are local coordinates of $\Sigma
_{t}$), $I_{0},...,I_{d-1}\in \left[ 0,..,d-1\right] $ denote internal
indices of the tensorial representation spaces of the Lorentz group, $e_{\mu
K}$ are the components of the co-tetrad one-form valued in the vectorial
representation space endowed with the flat metric $\eta
_{IJ}=diag(-1,1,...,1)$, $e=\det (e_{\mu I})$, and $e^{\mu K}$ is the
inverse of $e_{\mu L}$, $e^{\mu K}e_{\mu L}=\delta _{L}^{K}$, $e^{\mu
K}e_{\nu K}=\delta _{\nu }^{\mu }$. The metric $\eta _{IJ}$ and its inverse $%
\eta ^{IJ}$ are used to lower and to lift the Lorentz indices and to
determine the metric $g_{\mu \nu }=e_{\mu }^{I}e_{\nu }^{J}\eta _{IJ}$ of
the tangent space manifold $\mathcal{M}$ and $eA^{\mu K\nu L}=e(e^{\mu
K}e^{\nu L}-e^{\nu K}e^{\mu L})$. $\Psi $ is the Dirac spinors of mass $m$,
with components $\Psi _{A}$ where the Dirac indices take values $A,B\in
\left\{ 1,2,...,2(2^{\frac{d-2}{2}})\right\} $ for even dimension $d$ and $%
A,B\in \left\{ 1,2,...,2(2^{\frac{d-3}{2}})\right\} $ for old dimension $d$. 
$\overline{\Psi }=\Psi ^{\dag }\gamma ^{0}$ is the Dirac conjugate and $%
\gamma ^{I}$ are the Dirac Matrices\ satisfying%
\[
\gamma ^{K}\gamma ^{L}+\gamma ^{L}\gamma ^{K}=2\eta ^{KL}\text{.} 
\]%
The covariant derivative acts on the spinor fields as $D_{\mu }\Psi
=\partial _{\mu }\Psi +\omega _{1\mu KL}\frac{\sigma ^{KL}}{2}\Psi $ and $%
D_{\mu }\overline{\Psi }=\partial _{\mu }\overline{\Psi }-\omega _{1\mu KL}%
\overline{\Psi }\frac{\sigma ^{KL}}{2}$ where%
\[
\sigma ^{KL}=\frac{1}{4}(\gamma ^{K}\gamma ^{L}-\gamma ^{L}\gamma ^{K}) 
\]%
obeying The relations%
\begin{equation}
\left[ \sigma ^{KL},\sigma ^{PQ}\right] =\eta ^{KQ}\sigma ^{LP}+\eta
^{LP}\sigma ^{KQ}-\eta ^{KP}\sigma ^{LQ}+\eta ^{LQ}\sigma ^{KP}
\label{Lorentz-Gamma}
\end{equation}%
and%
\begin{equation}
\left[ \sigma ^{KL},\gamma _{N}\right] =\delta _{N}^{L}\gamma ^{K}-\delta
_{N}^{K}\gamma ^{L}\text{.}  \label{Transla-Gamma}
\end{equation}

\bigskip

The Lagrangian density of (\ref{LagrangianF}) is invariant under the
infinitesimal gauge transformations

\begin{eqnarray*}
\delta e_{\mu K} &=&\theta _{K}^{\text{ \ }N}e_{\mu N}\text{ \ , }\delta
\omega _{tNM}=-D_{t}\theta _{NM}\text{, \ }\delta \omega
_{1aKL}=-D_{1a}\theta _{KL}\text{,} \\
\delta \Psi &=&\frac{\sigma ^{KL}}{2}\Psi \theta _{KL}\text{ and }\delta 
\overline{\Psi }=-\overline{\Psi }\frac{\sigma ^{KL}}{2}\theta _{KL}
\end{eqnarray*}%
subject to the conditions \cite{Lagraa}%
\begin{equation}
D_{2a}\theta _{KL}=P_{2KaL}^{\text{ \ \ \ \ \ \ \ }PdQ}D_{a}\theta _{KL}=0%
\text{, and }\partial _{2a}\theta _{KL}=P_{2KaL}^{\text{ \ \ \ \ \ \ \ }%
PdQ}\partial _{a}\theta _{KL}=0\text{.}  \label{gauge-restrict}
\end{equation}

Note that these conditions do not restrict the gauge parameters $\theta
_{KL} $. They restrict only the gauge transformations of the dynamic part of
the connection.

The conjugate momenta $\pi ^{\beta N}$, $\mathcal{P}_{1}^{aKL}$ and $%
\mathcal{P}^{tKL}$ of the co-tetrad $e_{\beta N}$, $\omega _{1aKL}$ and $%
\omega _{tKL}$ are derived from the action (\ref{LagrangianF})%
\[
\pi ^{\beta N}(x)=\frac{\delta S}{\delta \partial _{t}e_{\beta N}(x)}=0\text{%
, }\mathcal{P}_{1}^{aKL}(x)=\frac{\delta S}{\delta \partial _{t}\omega
_{1aKL}(x)}=eA^{aKtL}(x) 
\]%
and 
\[
\mathcal{P}^{tKL}(x)=\frac{\delta S}{\delta \partial _{t}\omega _{tKL}(x)}=0%
\text{.} 
\]

The spinors $\Psi $ and $\overline{\Psi }$ are considered to be
anticommuting fields whose fermionic momenta $\Pi $ and $\overline{\Pi }$
are obtained from the left functional derivatives%
\[
\Pi =\frac{\delta S}{\delta \partial _{t}\overleftarrow{\Psi }(x)}=\frac{i}{2%
}ee^{tN}\overline{\Psi }\gamma _{N}\text{ and }\overline{\Pi }=\frac{\delta S%
}{\delta \partial _{t}\overleftarrow{\overline{\Psi }}(x)}=\frac{i}{2}%
ee^{tN}\gamma _{N}\Psi \text{.} 
\]

The bosonic phase space elements obey the following non-zero fundamental
Poisson brackets at fixed time%
\begin{eqnarray*}
\left\{ e_{\alpha I}(\overrightarrow{x}),\pi ^{\beta N}(\overrightarrow{y}%
)\right\} &=&\delta _{\alpha }^{\beta }\delta _{I}^{N}\delta (%
\overrightarrow{x}-\overrightarrow{y}), \\
\left\{ \omega _{tIJ}(\overrightarrow{x}),\mathcal{P}^{tKL}\left( 
\overrightarrow{y}\right) \right\} &=&\frac{1}{2}(\delta _{I}^{K}\delta
_{J}^{L}-\delta _{I}^{L}\delta _{J}^{K})\delta (\overrightarrow{x}-%
\overrightarrow{y}) \\
\left\{ \omega _{1aIJ}(\overrightarrow{x}),\mathcal{P}_{1}^{bKL}\left( 
\overrightarrow{y}\right) \right\} &=&P_{1IaJ}^{\text{ \ \ \ \ \ \ \ }%
KbL}\delta (\overrightarrow{x}-\overrightarrow{y})
\end{eqnarray*}%
where $\overrightarrow{x}$ denotes the local coordinates $x^{a}$ of $\Sigma
_{t}$. The fermionic ones obey the following non-zero fundamental
anticommuting Poisson brackets%
\[
\left\{ \Psi _{A}(\overrightarrow{x}),\Pi _{B}(\overrightarrow{y})\right\}
_{+}=\delta _{AB}\delta (\overrightarrow{x}-\overrightarrow{y})\text{ } 
\]%
and%
\[
\left\{ \overline{\Psi }_{A}(\overrightarrow{x}),\overline{\Pi }_{B}(%
\overrightarrow{y})\right\} _{+}=\delta _{AB}\delta (\overrightarrow{x}-%
\overrightarrow{y})\text{.} 
\]

The expressions of the conjugate momenta lead to the primary bosonic
constraints

\begin{eqnarray}
\pi ^{tN} &\simeq &0,\text{ }\mathcal{P}^{tKL}\simeq 0\text{ },  \nonumber \\
\pi ^{bN} &\simeq &0\text{, }C_{1}^{aKL}=\mathcal{P}_{1}^{aKL}-eA^{aKtL}%
\simeq 0  \label{PrimaryConst}
\end{eqnarray}%
and to the fermionic 
\begin{equation}
C=\Pi -\frac{i}{2}ee^{tN}\overline{\Psi }\gamma _{N}\text{, }\overline{C}=%
\overline{\Pi }-\frac{i}{2}ee^{tN}\gamma _{N}\Psi  \label{PrimaryFerconst}
\end{equation}%
satisfying the following non-zero Poisson brackets%
\begin{equation}
\left\{ \pi ^{aN}(\overrightarrow{x}),C_{1}^{bKL}(\overrightarrow{y}%
)\right\} =-eB^{aNtKbL}\delta (\overrightarrow{x}-\overrightarrow{y})\text{,}
\label{Pi-C1-Bracket}
\end{equation}%
\begin{equation}
\left\{ \pi ^{aN}(\overrightarrow{x}),C(\overrightarrow{y})\right\} =\frac{i%
}{2}eA^{aNtM}\overline{\Psi }(\overrightarrow{y})\gamma _{M}\delta (%
\overrightarrow{x}-\overrightarrow{y})\text{,}  \label{Pi-C-Bracket}
\end{equation}%
\begin{equation}
\left\{ \pi ^{aN}(\overrightarrow{x}),\overline{C}(\overrightarrow{y}%
)\right\} =\frac{i}{2}eA^{aNtM}\gamma _{M}\Psi (\overrightarrow{y})\delta (%
\overrightarrow{x}-\overrightarrow{y})  \label{Pi-Cbar-Bracket}
\end{equation}%
and 
\begin{equation}
\left\{ C_{A}(\overrightarrow{x}),\overline{C}_{B}(\overrightarrow{y}%
)\right\} _{+}=-iee^{tI}\gamma _{IBA}\delta (\overrightarrow{x}-%
\overrightarrow{y})\text{.}  \label{C-Cbar-Bracket}
\end{equation}%
\ \ 

The total Hamiltonian is%
\begin{equation}
\mathcal{H}_{T}=\int_{\Sigma }(\pi ^{tN}\Lambda _{tN}+\mathcal{P}^{tKL}\frac{%
\mathcal{A}_{tKL}}{2}+\pi ^{bN}\Lambda _{bN}+C_{1}^{aKL}\frac{\mathcal{A}%
_{1aKL}}{2}+C\Lambda -\overline{\Lambda }\overline{C})+H_{0}
\label{TOTALHAM}
\end{equation}%
where 
\begin{eqnarray*}
H_{0} &=&\int_{\Sigma }(eA^{aKbL}\frac{\Omega _{1abKL}}{2}%
+eA^{aKtL}D_{1a}\omega _{tKL}) \\
&&-\dint_{\Sigma }e\left( e^{aK}\frac{i}{2}\left( \overline{\Psi }\gamma
_{K}D_{a}\Psi -(D_{a}\overline{\Psi })\gamma _{K}\Psi \right) -m\overline{%
\Psi }\Psi \right) \\
&&-\dint_{\Sigma }ee^{tK}\frac{i}{2}\overline{\Psi }(\gamma _{K}\frac{\sigma
^{IJ}}{2}+\frac{\sigma ^{IJ}}{2}\gamma _{K})\Psi \omega _{tIJ}\text{,}
\end{eqnarray*}%
$\Lambda _{tN}$, $\mathcal{A}_{tKL}$, $\Lambda _{bN}$ and $\mathcal{A}%
_{1bKL} $ are the bosonic Lagrange multiplier fields and $\Lambda $ and $%
\overline{\Lambda }$ are the fermionic multiplier fields enforcing the
primary constraints.

The consistency requires that these primary constraints must be preserved
under the time evolution given by the total Hamiltonian (\ref{TOTALHAM}). In
order to satisfy the Jacobi identities, the calculation is given in terms of
Poisson brackets projected by $P_{1KaL}^{\text{ \ \ \ \ \ \ \ }PdQ}$ when
projected elements of the phase space $\omega _{1aKL}$ and $\mathcal{P}%
_{1}^{aKL}$ are involved \cite{Lagraa}:

\begin{eqnarray}
\left\{ \pi ^{tN},\mathcal{H}_{T}\right\} &=&-eB^{tNaKbL}\frac{\Omega
_{1abKL}}{2}+\frac{i}{2}eA^{tNaK}\left( \overline{\Psi }\gamma _{K}D_{a}\Psi
-(D_{a}\overline{\Psi })\gamma _{K}\Psi \right)  \nonumber \\
-ee^{tN}m\overline{\Psi }\Psi &=&P^{N}\simeq 0\text{,}  \label{PNFERM}
\end{eqnarray}

\begin{equation}
\left\{ \mathcal{P}^{tKL},\mathcal{H}_{T}\right\} =D_{a}eA^{aKtL}+\frac{i}{2}%
ee^{tI}\overline{\Psi }(\gamma _{I}\frac{\sigma ^{KL}}{2}+\frac{\sigma ^{KL}%
}{2}\gamma _{I})\Psi =M^{KL}\simeq 0,  \label{MNMFERM}
\end{equation}

\begin{eqnarray}
\left\{ \pi ^{bN},\mathcal{H}_{T}\right\} &=&-eB^{bNtKaL}\left( \frac{%
\mathcal{A}_{1aKL}}{2}-D_{1a}\omega _{tKL}\right) -eB^{bNaKcL}\frac{\Omega
_{1acKL}}{2}  \nonumber \\
&&+\frac{i}{2}eA^{bNtK}(\overline{\Psi }\gamma _{K}\Lambda -\overline{%
\Lambda }\gamma _{K}\Psi )  \nonumber \\
&&+\frac{i}{2}eA^{bNaK}\left( \overline{\Psi }\gamma _{K}D_{a}\Psi -(D_{a}%
\overline{\Psi })\gamma _{K}\Psi \right) -ee^{bN}m\overline{\Psi }\Psi 
\nonumber \\
&&+\frac{i}{2}eA^{bNtI}\left( \overline{\Psi }(\gamma _{I}\frac{\sigma ^{KL}%
}{2}+\frac{\sigma ^{KL}}{2}\gamma _{I})\Psi \right) \omega _{tKL}=\mathcal{R}%
^{bN}=0\text{,}  \label{EvolC1}
\end{eqnarray}%
\begin{eqnarray}
\left\{ C_{1}^{aKL},\mathcal{H}_{T}\right\} &=&eB^{bNtKaL}\left( \Lambda
_{bN}+\omega _{tN}^{\text{ \ \ }M}e_{bM}\right) +D_{1c}eA^{cKaL}  \nonumber
\\
&&+\frac{i}{2}(ee^{aI}\overline{\Psi }(\gamma _{I}\frac{\sigma ^{KL}}{2}+%
\frac{\sigma ^{KL}}{2}\gamma _{I})\Psi )_{1}=\mathcal{R}^{aKL}=0\text{,}
\label{EvolC2}
\end{eqnarray}

\begin{eqnarray}
\left\{ C,\mathcal{H}_{T}\right\} &=&\mathcal{C}_{fer}\simeq 0=-iee^{tN}%
\overline{\Lambda }\gamma _{N}-\frac{i}{2}eA^{bNtM}\Lambda _{bN}\overline{%
\Psi }\gamma _{M}-em\overline{\Psi }  \nonumber \\
&&-D_{a}(\frac{i}{2}ee^{aM}\overline{\Psi })\gamma _{M}-\frac{i}{2}%
ee^{aM}(D_{a}\overline{\Psi })\gamma _{M}  \nonumber \\
&&+\frac{i}{2}ee^{tM}\overline{\Psi }(\gamma _{M}\frac{\sigma ^{KL}}{2}+%
\frac{\sigma ^{KL}}{2}\gamma _{M})\omega _{tKL}  \label{EvolFerm}
\end{eqnarray}%
and 
\begin{eqnarray}
\left\{ \overline{C},\mathcal{H}_{T}\right\} &=&\overline{\mathcal{C}}%
_{fer}\simeq 0=-iee^{tN}\gamma _{N}\Lambda -\frac{i}{2}eA^{bNtM}\Lambda
_{bN}\gamma _{M}\Psi +em\Psi  \nonumber \\
&&-D_{a}(\frac{i}{2}ee^{aM}\gamma _{M}\Psi )-\frac{i}{2}ee^{aM}\gamma
_{M}(D_{a}\Psi )  \nonumber \\
&&-\frac{i}{2}ee^{tM}(\gamma _{M}\frac{\sigma ^{KL}}{2}+\frac{\sigma ^{KL}}{2%
}\gamma _{M})\Psi \omega _{tKL}\text{.}  \label{EvollFermbar}
\end{eqnarray}

These consistency conditions show that the evolution of the constraints $\pi
^{tN}$ and $\mathcal{P}^{tNM}$ lead to secondary constraints $P^{N}$ and $%
M^{NM}$ while the constraints $\pi ^{bN}$, $C_{1}^{aKL}$, $C$ and $\overline{%
C}$ lead to equations for Lagrange multipliers.

Now we have to check the consistency of the secondary constraints. For $%
M^{KL}$\ we combine (\ref{EvolC1}-\ref{EvollFermbar}), the properties (A.2)
of the B-matrix and the identity%
\begin{equation}
T_{\text{ }N}^{K}eA^{\nu N\rho L}+T_{\text{ }N}^{L}eA^{\nu K\rho N}=\frac{1}{%
2}\left( e_{\beta }^{K}eB^{\beta L\nu N\rho M}T_{NM}-e_{\beta }^{L}eB^{\beta
K\nu N\rho M}T_{NM}\right) \text{,}  \label{Identity-antis}
\end{equation}%
true for any antisymmetric tensor $T_{NM}=-T_{MN}$, to get%
\begin{eqnarray*}
&&\left\{ M^{KL},\mathcal{H}_{T}\right\} \\
&=&-\frac{1}{2}\left( e_{b}^{K}\mathcal{R}^{bL}-e_{b}^{L}\mathcal{R}%
^{bK}\right) +\frac{1}{2}(\overline{\Psi }\frac{\sigma ^{KL}}{2}\mathcal{C}%
_{fer}+\overline{\mathcal{C}}_{fer}\frac{\sigma ^{KL}}{2}\Psi ) \\
&&+\frac{1}{2}\left( e_{t}^{K}P^{L}-e_{t}^{L}P^{K}\right) -\left( \omega _{t%
\text{ }N}^{\text{ }K}M^{NL}+\omega _{t\text{ }N}^{\text{ }L}M^{KN}\right)
\end{eqnarray*}%
which, when (\ref{EvolC1}-\ref{EvollFermbar}) are satisfied, reduces to%
\begin{equation}
\left\{ M^{KL},\mathcal{H}_{T}\right\} =\frac{1}{2}\left(
e_{t}^{K}P^{L}-e_{t}^{L}P^{K}\right) -\left( \omega _{t\text{ }N}^{\text{ }%
K}M^{NL}+\omega _{t\text{ }N}^{\text{ }L}M^{KN}\right) \simeq 0
\label{evollorentz1}
\end{equation}%
ensuring the consistency of the constraint $M^{KL}$.

In what follows we consider instead of the constraint $P^{N}$ its temporal
projection%
\begin{equation}
\mathcal{D}_{t}=e_{tN}P^{N}=-eA^{aKbL}\frac{\Omega _{1abKL}}{2}+\frac{i}{2}%
ee^{aK}(\overline{\Psi }\gamma _{K}D_{a}\Psi -D_{a}\overline{\Psi }\gamma
_{K}\Psi )-em\overline{\Psi }\Psi  \label{Dtemp}
\end{equation}%
and its smeared spatial projection%
\begin{eqnarray}
\mathcal{D}_{sp}(\overrightarrow{N}) &=&-\int_{\Sigma
}N^{a}(e_{aN}P^{N}+\omega _{1aKL}M^{KL})  \nonumber \\
&=&\int_{\Sigma }eA^{aKtL}\mathcal{L}_{\overrightarrow{N}}(\omega _{1aKL})+%
\frac{i}{2}ee^{tK}(\overline{\Psi }\gamma _{K}\mathcal{L}_{\overrightarrow{N}%
}\Psi -\mathcal{L}_{\overrightarrow{N}}\overline{\Psi }\gamma _{K}\Psi )
\label{Diffeomorphism1}
\end{eqnarray}%
where $\mathcal{L}_{\overrightarrow{N}}(\omega _{1aKL})=N^{b}\partial
_{b}\omega _{1aKL}+\partial _{a}(N^{b})\omega _{1bKL}$ is the Lie-derivative
along the arbitrary $(d-1)$ dimensional vector field $\overrightarrow{N}$
tangent to $\Sigma $. This Lie-derivative $\mathcal{L}_{\overrightarrow{N}}$
treats the temporal components $e_{tN}$ and $\omega _{tKL}$ as well as the
Lorentz indices as scalars, i.e., $\mathcal{L}_{\overrightarrow{N}%
}(e_{tN})=N^{a}\partial _{a}e_{tN}$, $\mathcal{L}_{\overrightarrow{N}%
}(\omega _{tKL})=N^{a}\partial _{a}\omega _{tKL}$, $\mathcal{L}_{%
\overrightarrow{N}}\Psi =N^{b}\partial _{b}\Psi $ and $\mathcal{L}_{%
\overrightarrow{N}}\overline{\Psi }=N^{b}\partial _{b}\overline{\Psi }$ .

A straightforward computation gives for the evolution of the constraint $%
\mathcal{D}_{t}$ 
\begin{eqnarray*}
\left\{ \mathcal{D}_{t},\mathcal{H}_{T}\right\} &=&(\Lambda _{N}+\omega
_{tN}^{\text{ \ \ }M}e_{tM})P^{N}-D_{c}(e_{tN}\mathcal{R}^{cN})+\mathcal{R}%
^{aKL}(\frac{\mathcal{A}_{1aKL}}{2}-D_{a}\omega _{tKL}) \\
&&+(\Lambda _{cN}+\omega _{tN}^{\text{ \ \ }M}e_{cM})\mathcal{R}^{cN} \\
&&+\mathcal{C}_{fer}(\Lambda +\frac{\sigma ^{KL}}{2}\Psi \omega _{tKL})-(%
\overline{\Lambda }-\overline{\Psi }\frac{\sigma ^{KL}}{2}\omega _{tKL})%
\overline{\mathcal{C}}_{fer}
\end{eqnarray*}%
leading to%
\begin{equation}
\left\{ \mathcal{D}_{t},\mathcal{H}_{T}\right\} =\left( \Lambda _{N}+\omega
_{tN}^{\text{ \ \ }M}e_{tM}\right) P^{N}\simeq 0  \label{DetLamdaN}
\end{equation}%
when (\ref{EvolC1} -\ref{EvollFermbar}) are satisfied. For%
\begin{equation}
\Lambda _{N}=-\omega _{tN}^{\text{ \ \ }M}e_{tM}\Longrightarrow \pi
^{tN}\Lambda _{N}=-\frac{1}{2}(\pi ^{N}e_{t}^{M}-\pi ^{M}e_{t}^{N})\omega
_{tNM}  \label{DetLamda1}
\end{equation}%
(\ref{DetLamdaN}) vanishes strongly. The consistency of $\mathcal{D}_{sp}(%
\overrightarrow{N})$ gives%
\begin{eqnarray*}
\left\{ \mathcal{D}_{sp}(\overrightarrow{N}),\mathcal{H}_{T}\right\}
&=&-\int_{\Sigma }(P^{N}\mathcal{L}_{\overrightarrow{N}}(e_{tN})+M^{KL}%
\mathcal{L}_{\overrightarrow{N}}(\omega _{tKL}))-\mathcal{R}^{aKL}\mathcal{L}%
_{\overrightarrow{N}}(\omega _{1aKL}) \\
&&-\mathcal{R}^{cN}\mathcal{L}_{\overrightarrow{N}}(e_{cN})-\mathcal{C}_{fer}%
\mathcal{L}_{\overrightarrow{N}}\Psi +\mathcal{L}_{\overrightarrow{N}}(%
\overline{\Psi })\overline{\mathcal{C}}_{fer}
\end{eqnarray*}%
leading to%
\[
\left\{ \mathcal{D}_{sp}(\overrightarrow{N}),\mathcal{H}_{T}\right\}
=-\int_{\Sigma }(P^{N}\mathcal{L}_{\overrightarrow{N}}(e_{tN})+M^{KL}%
\mathcal{L}_{\overrightarrow{N}}(\omega _{tKL})) 
\]%
when (\ref{EvolC1}-\ref{EvollFermbar}) are satisfied.

This shows that the set of constrains is complete meaning that the total
Hamiltonian $\mathcal{H}_{T}$ is consistent providing that (\ref{EvolC1} -%
\ref{EvollFermbar}) are satisfied.

Now we have to solve equations (\ref{EvolC1}-\ref{EvollFermbar}) to
determine the Lagrange multipliers and then insert their expressions into
the Hamiltonian. As consequence of \ $\mathcal{R}^{dPQ}=P_{1KaL}^{\text{ \ \
\ \ \ \ \ }PdQ}\mathcal{R}^{aKL}$ and the rank $d(d-1)$ of the projector $%
P_{1KaL}^{\text{ \ \ \ \ \ \ \ }PdQ}$, there are as many equations as
multipliers of Lagrange $\Lambda _{bN}$. This is an indication to uniquely
solve (\ref{EvolC2}) by multiplying it by $B_{bNtKaL}$ and using (A.6) to get%
\begin{eqnarray}
\Lambda _{bN} &=&-\omega _{tN}^{\text{ \ \ }%
M}e_{bM}+D_{b}e_{tN}-e^{-1}B_{bNtKtL}M^{KL}  \nonumber \\
&&-\frac{i}{4}A_{bKtL}\overline{\Psi }(\gamma _{N}\frac{\sigma ^{KL}}{2}+%
\frac{\sigma ^{KL}}{2}\gamma _{N})\Psi  \nonumber \\
&\simeq &-\omega _{tN}^{\text{ \ \ }M}e_{bM}+D_{b}e_{tN}-\frac{i}{4}A_{bKtL}%
\overline{\Psi }(\gamma _{N}\frac{\sigma ^{KL}}{2}+\frac{\sigma ^{KL}}{2}%
\gamma _{N})\Psi \text{.}  \label{MultiabN}
\end{eqnarray}

Similarly for $\mathcal{A}_{1aKL}$ which has the same number of components
as equations (\ref{EvolC1}) whose solution is%
\begin{eqnarray}
\frac{1}{2}\mathcal{A}_{1aKL} &=&D_{1a}\omega _{tKL}-B_{bNtKaL}B^{bNcPdQ}%
\frac{\Omega _{1cdPQ}}{2}  \label{MultAaKL1} \\
&&+\frac{i}{2}B_{bNtKaL}A^{bNtM}(\overline{\Psi }\gamma _{M}\Lambda -%
\overline{\Lambda }\gamma _{M}\Psi )-B_{bNtKaL}e^{bN}m\overline{\Psi }\Psi 
\nonumber \\
&&+\frac{i}{2}B_{bNtKaL}A^{bNcM}(\overline{\Psi }\gamma _{M}D_{c}\Psi -D_{c}%
\overline{\Psi }\gamma _{M}\Psi )  \nonumber \\
&&+\frac{i}{2}B_{bNtKaL}A^{bNtM}\overline{\Psi }(\gamma _{M}\frac{\sigma
^{PQ}}{2}+\frac{\sigma ^{PQ}}{2}\gamma _{M})\Psi \omega _{tPQ}\text{.} 
\nonumber
\end{eqnarray}

By multiplying on the left (\ref{EvolFerm}) and on the right (\ref%
{EvollFermbar}) by $i\frac{e^{-1}}{g^{tt}}e^{tM}\gamma _{M}$ and using%
\begin{eqnarray*}
-\frac{i}{2}eA^{bNtM}\Lambda _{bN} &=&\frac{i}{2}e\omega _{t\text{ }%
N}^{M}e^{tN}+D_{b}(\frac{i}{2}ee^{bM})+\frac{i}{2}M^{NM}e_{tN} \\
&\simeq &\frac{i}{2}e\omega _{t\text{ }N}^{M}e^{tN}+D_{b}(\frac{i}{2}ee^{bM})
\end{eqnarray*}%
deduced from (\ref{MultiabN}), we obtain the expressions of the fermionic
Lagrange multipliers%
\begin{equation}
\Lambda =-\frac{\sigma ^{KL}}{2}\Psi \omega _{tKL}-\frac{i}{g^{tt}}%
me^{tM}\gamma _{M}\Psi -\frac{1}{g^{tt}}e^{bK}e^{tM}\gamma _{M}\gamma
_{K}D_{b}\Psi  \label{Multfermion}
\end{equation}%
and 
\begin{equation}
\overline{\Lambda }=\overline{\Psi }\frac{\sigma ^{KL}}{2}\omega _{tKL}+%
\frac{i}{g^{tt}}me^{tM}\overline{\Psi }\gamma _{M}-\frac{1}{g^{tt}}%
e^{bK}e^{tM}D_{b}\overline{\Psi }\gamma _{K}\gamma _{M}
\label{multifermionBar}
\end{equation}%
where $g^{tt}=e^{tN}e_{N}^{t}$. The Insertion of (\ref{Multfermion}-\ref%
{multifermionBar}) in (\ref{MultAaKL1}) gives%
\begin{eqnarray*}
\frac{1}{2}\mathcal{A}_{1aKL} &=&D_{1a}\omega _{tKL}-B_{bNtKaL}(B^{bNcPdQ}%
\frac{\Omega _{1cdPQ}}{2} \\
&&-\frac{i}{2}A^{bNcM}(\overline{\Psi }\gamma _{M}D_{c}\Psi -D_{c}\overline{%
\Psi }\gamma _{M}\Psi ) \\
&&+\frac{i}{2g^{tt}}A^{bNtP}e^{cM}e^{tI}(\overline{\Psi }\gamma _{P}\gamma
_{I}\gamma _{M}D_{c}\Psi -D_{c}\overline{\Psi }\gamma _{M}\gamma _{I}\gamma
_{P}\Psi ) \\
&&+\frac{m}{g^{tt}}e^{bP}e_{P}^{t}e^{tN}\overline{\Psi }\Psi )\text{.}
\end{eqnarray*}%
which is inserted with (\ref{DetLamda1}), (\ref{MultiabN}) and (\ref%
{Multfermion}-\ref{multifermionBar}) in (\ref{TOTALHAM}) to give the total
Hamiltonian 
\begin{eqnarray}
\mathcal{H}_{T}^{\prime } &=&\int_{\Sigma }\mathcal{P}^{tKL}\frac{\mathcal{A}%
_{tKL}}{2}-\int_{\Sigma }(\frac{1}{2}(\pi ^{\mu K}e_{\mu }^{L}-\pi ^{\mu
L}e_{\mu }^{K})+D_{a}(C_{1}^{aKL}+eA^{aKtL})  \nonumber \\
&&+(C+\frac{i}{2}ee^{tI}\overline{\Psi }\gamma _{I})\frac{\sigma ^{KL}}{2}%
\Psi +\overline{\Psi }\frac{\sigma ^{KL}}{2}(\overline{C}+\frac{i}{2}%
ee^{tI}\gamma _{I}\Psi ))\omega _{tKL}  \nonumber \\
&&+\int_{\Sigma }(eA^{aKbL}\frac{\Omega _{1abKL}}{2}-\frac{i}{2}ee^{aK}(%
\overline{\Psi }\gamma _{K}D_{a}\Psi -D_{a}\overline{\Psi }\gamma _{K}\Psi
)+em\overline{\Psi }\Psi )  \nonumber \\
&&+\int_{\Sigma }\pi ^{aN}\left( D_{a}e_{tN}-\frac{i}{4}A_{aKtL}\overline{%
\Psi }(\gamma _{N}\frac{\sigma ^{KL}}{2}+\frac{\sigma ^{KL}}{2}\gamma
_{N})\Psi \right)  \nonumber \\
&&-\int_{\Sigma }C_{1}^{aKL}B_{bNtKaL}(B^{bNcPdQ}\frac{\Omega _{cdPQ}}{2}-%
\frac{i}{2}A^{bNcM}(\overline{\Psi }\gamma _{M}D_{c}\Psi -D_{c}\overline{%
\Psi }\gamma _{M}\Psi )  \nonumber \\
&&+\frac{i}{2g^{tt}}e^{tP}e^{cI}A^{bNtM}(\overline{\Psi }\gamma _{M}\gamma
_{P}\gamma _{I}D_{c}\Psi -D_{c}\overline{\Psi }\gamma _{I}\gamma _{P}\gamma
_{M}\Psi )  \nonumber \\
&&+\frac{1}{g^{tt}}e_{M}^{t}e^{bM}e^{tN}m\overline{\Psi }\Psi )-\int_{\Sigma
}m\frac{ie^{tM}}{g^{tt}}(C\gamma _{M}\Psi +\overline{\Psi }\gamma _{M}%
\overline{C})  \nonumber \\
&&-\int_{\Sigma }\frac{1}{g^{tt}}e^{tM}e^{aI}(C\gamma _{M}\gamma
_{I}D_{a}\Psi -D_{a}\overline{\Psi }\gamma _{I}\gamma _{M}\overline{C})\text{%
.}  \label{hAMILTON1}
\end{eqnarray}

With the Hamiltonian $\mathcal{H}_{T}^{\prime }$, the consistency of the
constraint $\mathcal{P}^{tNM}$ leads to%
\begin{eqnarray}
\left\{ \mathcal{P}^{tKL},\mathcal{H}_{T}^{\prime }\right\}
&=&D_{a}(C_{1}^{aKL}+eA^{aKtL})+\frac{1}{2}(\pi ^{\mu K}e_{\mu }^{L}-\pi
^{\mu L}e_{\mu }^{K}))  \nonumber \\
&&+(C+\frac{i}{2}e^{tI}\overline{\Psi }\gamma _{I})\frac{\sigma ^{KL}}{2}%
\Psi +\overline{\Psi }\frac{\sigma ^{KL}}{2}(\overline{C}+\frac{i}{2}%
e^{tI}\gamma _{I}\Psi )  \nonumber \\
&=&\frac{1}{2}M^{\prime KL}\simeq 0  \label{MKL1}
\end{eqnarray}%
where $M^{\prime KL}$ is the new constraints replacing $M^{KL}$. \ By using
the relations $e_{tN}\frac{\delta }{\delta e_{tN}}(B_{bMtKaL})=B_{bMtKaL}$, $%
e_{tN}\left\{ \pi ^{tN},\frac{e^{tM}}{g^{tt}}\right\} =-\frac{e^{tM}}{g^{tt}}
$ and $e_{tN}\left\{ \pi ^{tN},e^{bM}\right\} =0$ we can rewrite the total
Hamiltonian (\ref{hAMILTON1}) as%
\[
\mathcal{H}_{T}^{\prime }=\int_{\Sigma }(\frac{1}{2}\mathcal{P}^{tKL}%
\mathcal{A}_{tKL}-\mathcal{D}_{t}^{\prime }-M^{\prime KL}\frac{\omega _{tKL}%
}{2}) 
\]%
where $\mathcal{D}_{t\text{ }}^{\prime }$ is the projection of the evolution
of $\pi ^{tN}$, i.e.,%
\begin{eqnarray*}
e_{tN}\left\{ \pi ^{tN},\mathcal{H}_{T}^{\prime }\right\} &=&-e_{tN}\omega
_{tM}^{N}\pi ^{tM}+P^{\prime N}e_{tN} \\
&=&-e_{tN}\omega _{tM}^{N}\pi ^{tM}+\mathcal{D}_{t}^{\prime }\simeq
0\Longrightarrow \mathcal{D}_{t}^{\prime }\simeq 0.
\end{eqnarray*}

By using $e_{cN}\frac{\delta }{\delta e_{tN}}%
(e^{-1}B_{bMtKaL})=e^{-1}B_{bMcKaL}$, $C_{1}^{aKL}B_{bMcKaL}=0$ and%
\begin{eqnarray*}
&&\dint_{\Sigma }C_{1}^{aKL}B_{bMtKaL}N^{c}e_{cN}C^{tNbMePdQ}\frac{\Omega
_{1edPQ}}{2} \\
&=&-\dint_{\Sigma }(C_{1}^{aKL}\mathcal{L}_{\overrightarrow{N}}(\omega
_{1aKL})+D_{a}(eA^{aKtL})N^{c}\omega _{1cKL}) \\
&&-\dint_{\Sigma }N^{c}C_{1}^{aKL}e^{-1}B_{cMtKaL}eB^{tMePdQ}\frac{\Omega
_{1edPQ}}{2}
\end{eqnarray*}%
we get for the projection $e_{cN}P^{\prime N}$, the following smeared
combination of constraints%
\begin{eqnarray*}
&&\dint_{\Sigma }N^{c}(e_{cN}P^{\prime N}+\frac{\omega _{cKL}}{2}M^{\prime
KL}+\pi ^{aN}(D_{a}e_{cN}-D_{c}e_{aN})) \\
&&-\frac{i}{4}\dint_{\Sigma }N^{c}\pi ^{aN}A_{aKcL}\overline{\Psi }(\gamma
_{N}\frac{\sigma ^{KL}}{2}+\frac{\sigma ^{KL}}{2}\gamma _{N})\Psi
-\dint_{\Sigma }N^{c}C_{1}^{aKL}e^{-1}B_{cMtKaL}P^{M} \\
&=&-\dint_{\Sigma }(\pi ^{aM}\mathcal{L}_{\overrightarrow{N}%
}(e_{aM})+(C_{1}^{aKL}+eA^{aKtL})\mathcal{L}_{\overrightarrow{N}}(\omega
_{1aKL})) \\
&&.-\dint_{\Sigma }((C+\frac{i}{2}ee^{tN}\overline{\Psi }\gamma _{N})%
\mathcal{L}_{\overrightarrow{N}}(\Psi )-\mathcal{L}_{\overrightarrow{N}}(%
\overline{\Psi })(\overline{C}+\frac{i}{2}ee^{tN}\gamma _{N}\Psi )).
\end{eqnarray*}

This is completed by adding the constraint $\pi ^{tN}\mathcal{L}_{%
\overrightarrow{N}}(e_{tN})$ to get 
\begin{eqnarray*}
\mathcal{D}_{sp}^{\prime }(\overrightarrow{N}) &=&\dint_{\Sigma }(\pi ^{\mu
M}\mathcal{L}_{\overrightarrow{N}}(e_{\mu M})+(C_{1}^{aKL}+eA^{aKtL})%
\mathcal{L}_{\overrightarrow{N}}(\omega _{1aKL})) \\
&&+\dint_{\Sigma }((C+\frac{i}{2}ee^{tN}\overline{\Psi }\gamma _{N})\mathcal{%
L}_{\overrightarrow{N}}(\Psi )-\mathcal{L}_{\overrightarrow{N}}(\overline{%
\Psi })(\overline{C}+\frac{i}{2}ee^{tN}\gamma _{N}\Psi )).
\end{eqnarray*}

A straightforward computation shows that $\mathcal{D}_{sp}^{\prime }(%
\overrightarrow{N})$\ satisfies 
\[
\left\{ \mathcal{D}_{sp}^{\prime }(\overrightarrow{N}),\mathcal{D}%
_{sp}^{\prime }(\overrightarrow{N^{\prime }})\right\} =\mathcal{D}%
_{sp}^{\prime }(\mathcal{L}_{\overrightarrow{N}}(\overrightarrow{N^{\prime }}%
)-\mathcal{L}_{\overrightarrow{N^{\prime }}}(\overrightarrow{N}))=\mathcal{D}%
_{sp}^{\prime }(\left[ \overrightarrow{N},\overrightarrow{N^{\prime }}\right]
)\text{,} 
\]%
where$\left[ \overrightarrow{N},\overrightarrow{N^{\prime }}\right] $ is the
Lie bracket. The constraint $\mathcal{D}_{sp}^{\prime }(\overrightarrow{N})$%
\ generates\ spatial diffeomorphisms of the phase space elements 
\[
\left\{ e_{\mu N},\mathcal{D}_{sp}^{\prime }(\overrightarrow{N})\right\} =%
\mathcal{L}_{\overrightarrow{N}}(e_{\mu N})\text{, }\left\{ \omega _{1aNM},%
\mathcal{D}_{sp}^{\prime }(\overrightarrow{N})\right\} =\mathcal{L}_{%
\overrightarrow{N}}(\omega _{1aNM})\text{,} 
\]%
\[
\left\{ \Psi ,\mathcal{D}_{sp}^{\prime }(\overrightarrow{N})\right\} =%
\mathcal{L}_{\overrightarrow{N}}(\Psi )\text{, }\left\{ \overline{\Psi },%
\mathcal{D}_{sp}^{\prime }(\overrightarrow{N})\right\} =\mathcal{L}_{%
\overrightarrow{N}}(\overline{\Psi })\text{,} 
\]%
\[
\left\{ \pi ^{\mu N},\mathcal{D}_{sp}^{\prime }(\overrightarrow{N})\right\} =%
\mathcal{L}_{\overrightarrow{N}}(\pi ^{\mu N})\text{, }\left\{ C^{aNM},%
\mathcal{D}_{sp}^{\prime }(\overrightarrow{N})\right\} =\mathcal{L}_{%
\overrightarrow{N}}(C^{aNM})\text{,} 
\]%
\begin{equation}
\left\{ C,\mathcal{D}_{sp}^{\prime }(\overrightarrow{N})\right\} =\mathcal{L}%
_{\overrightarrow{N}}(C)\text{ and }\left\{ \overline{C},\mathcal{D}%
_{sp}^{\prime }(\overrightarrow{N})\right\} =\mathcal{L}_{\overrightarrow{N}%
}(\overline{C})  \label{Constr-Trans-Diff}
\end{equation}%
from which we deduce that the Poisson brackets of $\mathcal{D}_{sp}^{\prime
}(N)$ with the primary constraints weakly vanish and the constraints $%
\mathcal{D}_{t}^{\prime }$ and $M^{\prime KL}$ transform like scalar
densities of weight one 
\begin{eqnarray*}
\left\{ \mathcal{D}_{t}^{\prime },\mathcal{D}_{sp}^{\prime }(\overrightarrow{%
N})\right\} &=&\mathcal{L}_{\overrightarrow{N}}(\mathcal{D}_{t}^{\prime
})=\partial _{c}(N^{c}\mathcal{D}_{t}^{\prime }) \\
&\Longrightarrow &\left\{ \mathcal{D}_{sp}^{\prime }(\overrightarrow{N}),%
\mathcal{D}_{t}^{\prime }(M)\right\} =\mathcal{D}_{t}^{\prime }(\mathcal{L}_{%
\overrightarrow{N}}(M))
\end{eqnarray*}%
and%
\begin{eqnarray*}
\left\{ M^{KL},\mathcal{D}_{sp}^{\prime }(\overrightarrow{N})\right\} &=&%
\mathcal{L}_{\overrightarrow{N}}(M^{KL})=\partial _{c}(N^{c}M^{KL}) \\
&\Longrightarrow &\left\{ \mathcal{D}_{sp}^{\prime }(\overrightarrow{N}),%
\mathcal{M}(\theta )\right\} =\mathcal{M}(\mathcal{L}_{\overrightarrow{N}%
}(\theta ))\text{.}
\end{eqnarray*}

Here $\mathcal{D}_{t}^{\prime }(M)=\dint_{\Sigma }M\mathcal{D}_{t}^{\prime }$
and $\mathcal{M}(\theta )=\dint_{\Sigma }M^{\prime KL}\frac{\theta _{KL}}{2}$
where $\theta _{KL}=\delta t\omega _{tKL}$ are dimensionless and
infinitesimal arbitrary local parameters subject to the condition (\ref%
{gauge-restrict}). The above Poisson brackets imply that the constraint $%
\mathcal{D}_{sp}^{\prime }(\overrightarrow{N})$ is of first-class.

For the smeared constraint $\mathcal{M}(\theta )$ which acts\ as generators
of the infinitesimal Lorentz transformation group, we get%
\begin{equation}
\left\{ e_{\mu N},\mathcal{M}(\theta )\right\} =\theta _{N}^{\text{ \ \ }%
L}e_{\mu L}\text{, }\left\{ \omega _{1aNM},\mathcal{M}(\theta )\right\} =-%
\mathcal{D}_{1a}(\theta _{NM})\text{,}  \label{PH-SP6LTR1}
\end{equation}%
\begin{equation}
\left\{ \Psi ,\mathcal{M}(\theta )\right\} =\frac{\sigma ^{KL}}{2}\Psi
\theta _{KL}\text{, }\left\{ \overline{\Psi },\mathcal{M}(\theta )\right\} =-%
\overline{\Psi }\frac{\sigma ^{KL}}{2}\theta _{KL}\text{.}
\label{Psi-SPLTR1}
\end{equation}

\begin{equation}
\left\{ \pi ^{\mu N},\mathcal{M}(\theta )\right\} =\theta _{\text{ \ }%
L}^{N}\pi ^{\mu L}\text{, }\left\{ C_{1}^{aNM},\mathcal{M}(\theta )\right\}
=\theta _{\text{ \ }L}^{N}C_{1}^{aLM}+\theta _{\text{ \ }L}^{M}C_{1}^{aNL}
\label{PH-SPLTR2}
\end{equation}%
\begin{equation}
\left\{ C,\mathcal{M}(\theta )\right\} =-C\frac{\sigma ^{KL}}{2}\theta _{KL}%
\text{, }\left\{ \overline{C},\mathcal{M}(\theta )\right\} =\frac{\sigma
^{KL}}{2}\overline{C}\theta _{KL}  \label{C-SPLTR2}
\end{equation}%
from which we deduce that the Poisson brackets of $\mathcal{M}(\theta )$
with the primary constraints $\pi ^{aN}$, $C^{aNM}$, $C$ and $\overline{C}$
weakly vanish. Since\ $\mathcal{M}(\theta )$ treats the space-time indices
as scalars, the transformations (\ref{PH-SP6LTR1}-\ref{C-SPLTR2}) make easy
the computation of transformation it generates. the constraints $M^{^{\prime
}KL}$ transform as tensor%
\[
\left\{ M^{\prime NM},\mathcal{M}(\theta )\right\} =\theta _{\text{ \ }%
L}^{N}M^{\prime LM}+\theta _{\text{ \ }L}^{M}M^{\prime NL}\text{,} 
\]%
leading to the $so(1,d-1)$ Lie algebra%
\begin{eqnarray*}
\left\{ M^{\prime NM}(\overrightarrow{x}),M^{\prime KL}(\overrightarrow{y}%
)\right\} _{D} &=&(\eta ^{NL}M^{\prime MK}(\overrightarrow{x})+\eta
^{MK}M^{\prime NL}(\overrightarrow{x}) \\
&&-\eta ^{NK}M^{\prime ML}(\overrightarrow{x})-\eta ^{ML}M^{\prime NK}(%
\overrightarrow{x}))\delta (\overrightarrow{x}-\overrightarrow{y})\text{,}
\end{eqnarray*}%
and $\mathcal{D}_{t}^{\prime }$ is a scalar under Lorentz transformations,
implying%
\[
\left\{ \mathcal{D}_{t}^{\prime }(\overrightarrow{x}),\mathcal{M}(\theta
)\right\} =0\Longrightarrow \left\{ \mathcal{D}_{t}^{\prime }(M),\mathcal{M}%
(\theta )\right\} =0 
\]%
showing that the constraints $M^{\prime KL}$ are also of first-class.

Now, what remains is to study the consistency of the constraint $\mathcal{D}%
_{t}^{\prime }$. Since $\mathcal{D}_{t}^{\prime }$ commutes weakly in terms
of Poisson brackets with $\mathcal{D}_{sp}^{\prime }(\overrightarrow{N})$
and $\mathcal{M}(\theta )$, it remains to calculate its Poisson brackets
with the primary constraints $\pi ^{aN}$, $C_{1}^{aNM}$, $C$ and $\overline{C%
}$.\ A straightforward computation gives%
\begin{eqnarray*}
&&\left\{ \pi ^{cN}(\overrightarrow{x}),\mathcal{D}_{t}^{\prime }(%
\overrightarrow{y})\right\} \simeq +(eB^{cNaKbL}\frac{\Omega _{1abKL}}{2} \\
&&-\frac{i}{2}eA^{cNaK}(\overline{\Psi }\gamma _{K}D_{a}\Psi -D_{a}\overline{%
\Psi }\gamma _{K}\Psi )+ee^{cN}\overline{\Psi }\Psi )\delta (\overrightarrow{%
x}-\overrightarrow{y}) \\
&&-eB^{cNtKaL}B_{bMtKaL}(B^{bMePdQ}\frac{\Omega _{1edPQ}}{2}-\frac{i}{2}%
A^{bMdP}(\overline{\Psi }\gamma _{P}D_{d}\Psi -D_{d}\overline{\Psi }\gamma
_{P}\Psi ) \\
&&+\frac{i}{2g^{tt}}e^{tP}e^{dI}A^{bMtQ}(\overline{\Psi }\gamma _{Q}\gamma
_{P}\gamma _{I}D_{d}\Psi -D_{d}\overline{\Psi }\gamma _{I}\gamma _{P}\gamma
_{Q}\Psi ) \\
&&+m\frac{1}{g^{tt}}e_{I}^{t}e^{bI}e^{tM}\overline{\Psi }\Psi ))\delta (%
\overrightarrow{x}-\overrightarrow{y})-m\frac{1}{g^{tt}}eA^{cNtI}e_{I}^{t}%
\overline{\Psi }\Psi \delta (\overrightarrow{x}-\overrightarrow{y}) \\
&&+\frac{i}{2g^{tt}}e^{tP}e^{dI}eA^{cNtQ}(\overline{\Psi }\gamma _{Q}\gamma
_{P}\gamma _{I}D_{d}\Psi -D_{d}\overline{\Psi }\gamma _{I}\gamma _{P}\gamma
_{Q}\Psi )\delta (\overrightarrow{x}-\overrightarrow{y})
\end{eqnarray*}%
where $\simeq $ means that only the terms which are not proportional to the
primary constraints have been kept. $\left\{ \pi ^{cN}(\overrightarrow{x}),%
\mathcal{D}_{t}^{\prime }(\overrightarrow{y})\right\} $ weakly vanishes as a
consequence of (A.6) applied in the third line.

For the Poisson brackets of the constraint $C_{1}^{aNM}$ with $\mathcal{D}%
_{t}^{f}$ we get 
\begin{eqnarray*}
&&\left\{ C_{1}^{aKL}(\overrightarrow{x}),\mathcal{D}_{t}^{f}(%
\overrightarrow{y})\right\} \\
&\simeq &-(D_{b}eA^{bKaL})_{1}\delta (\overrightarrow{x}-\overrightarrow{y})
\\
&&-\frac{i}{2}(e^{aN}(\overline{\Psi }\gamma _{N}\frac{\sigma ^{KL}}{2}+%
\frac{\sigma ^{KL}}{2}\gamma _{N}\Psi ))_{1}\delta (\overrightarrow{x}-%
\overrightarrow{y}) \\
&&-eB^{cNtKaL}D_{c}e_{tN}\delta (\overrightarrow{x}-\overrightarrow{y}) \\
&&+\frac{i}{4}eB^{cNtKaL}A_{cPtQ}(\overline{\Psi }\gamma _{N}\frac{\sigma
^{PQ}}{2}+\frac{\sigma ^{PQ}}{2}\gamma _{N}\Psi ))\delta (\overrightarrow{x}-%
\overrightarrow{y})\text{.}
\end{eqnarray*}

The first term of the second hand gives%
\begin{eqnarray*}
&&D_{1b}eA^{bKaL} \\
&=&B_{cMtPdQ}B^{cMtKal}D_{b}eA^{bPdQ} \\
&=&B_{cMtPdQ}B^{cMtKal}(eB^{tIbPdQ}D_{b}e_{tI}+eB^{eIbPdQ}D_{b}e_{eI}) \\
&=&-eB^{cMtKal}D_{c}e_{tM}-B_{cMtPtQ}B^{cMtKal}D_{b}eA^{bPtQ}\text{.}
\end{eqnarray*}

The relation%
\[
B_{cMtPdQ}e^{dN}=(-B_{cMtPtQ}e^{tN}+\frac{1}{2}(\frac{A_{cMtP}}{d-2}\delta
_{Q}^{N}+\frac{A_{cQtM}}{d-2}\delta _{P}^{N}+A_{cPtQ}\delta _{M}^{N})\text{,}
\]%
deduced from (A.3), and 
\[
\frac{1}{2}(\frac{A_{cMtP}}{d-2}\delta _{Q}^{N}+\frac{A_{cQtM}}{d-2}\delta
_{P}^{N})(\overline{\Psi }\gamma _{N}\frac{\sigma ^{PQ}}{2}+\frac{\sigma
^{PQ}}{2}\gamma _{N}\Psi )=0 
\]%
imply for the second term 
\begin{eqnarray*}
&&-\frac{i}{2}(e^{aN}(\overline{\Psi }\gamma _{N}\frac{\sigma ^{KL}}{2}+%
\frac{\sigma ^{KL}}{2}\gamma _{N}\Psi ))_{1} \\
&=&-\frac{i}{2}B_{cMtPdQ}B^{cMtKal}e^{dN}(\overline{\Psi }\gamma _{N}\frac{%
\sigma ^{PQ}}{2}+\frac{\sigma ^{PQ}}{2}\gamma _{N}\Psi ) \\
&=&\frac{i}{2}B_{cMtPtQ}B^{cMtKal}e^{tN}(\overline{\Psi }\gamma _{N}\frac{%
\sigma ^{PQ}}{2}+\frac{\sigma ^{PQ}}{2}\gamma _{N}\Psi ) \\
&&-\frac{i}{4}eB^{cMtKaL}A_{cPtQ}(\overline{\Psi }\gamma _{M}\frac{\sigma
^{PQ}}{2}+\frac{\sigma ^{PQ}}{2}\gamma _{M}\Psi )
\end{eqnarray*}%
leading to%
\begin{eqnarray*}
&&\left\{ C_{1}^{aKL}(\overrightarrow{x}),\mathcal{D}_{t}^{\prime }(%
\overrightarrow{y})\right\} \\
&\simeq &B_{cMtPtQ}B^{cMtKal}(D_{b}eA^{bPtQ}+\frac{i}{2}(e^{tN}(\overline{%
\Psi }\gamma _{N}\frac{\sigma ^{PQ}}{2}+\frac{\sigma ^{PQ}}{2}\gamma
_{N}\Psi ))\delta (\overrightarrow{x}-\overrightarrow{y}) \\
&\simeq &B_{cMtPtQ}B^{cMtKal}M^{KL}\delta (\overrightarrow{x}-%
\overrightarrow{y})\simeq 0\text{.}
\end{eqnarray*}

For the fermionic constraints, we get%
\begin{eqnarray*}
\left\{ C(\overrightarrow{x}),\mathcal{D}_{t}^{\prime }(\overrightarrow{y}%
)\right\} &\simeq &((D_{a}(\frac{i}{2}e^{aK}\overline{\Psi })+\frac{i}{2}%
e^{aK}D_{a}\overline{\Psi })\gamma _{K}+em\overline{\Psi })\delta (%
\overrightarrow{x}-\overrightarrow{y}) \\
&&\frac{i}{2}eA^{cNtK}\overline{\Psi }\gamma _{K}(D_{c}e_{tN} \\
&&-\frac{i}{4}A_{cPtQ}\overline{\Psi }(\gamma _{N}\frac{\sigma ^{PQ}}{2}+%
\frac{\sigma ^{PQ}}{2}\gamma _{N})\Psi )\delta (\overrightarrow{x}-%
\overrightarrow{y}) \\
&&-\frac{im}{g^{tt}}e^{tM}\overline{\Psi }(\overrightarrow{y})\gamma
_{M}\left\{ C(\overrightarrow{x}),\overline{C}(\overrightarrow{y})\right\} \\
&&+\frac{1}{g^{tt}}e^{tM}e^{aK}D_{a}\overline{\Psi }(\overrightarrow{y}%
)\gamma _{K}\gamma _{M}\left\{ C(\overrightarrow{x}),\overline{C}(%
\overrightarrow{y})\right\} \text{.}
\end{eqnarray*}

The relations

\begin{eqnarray*}
eA^{cNtK}D_{c}e_{tN} &=&D_{c}(eA^{cNtK}e_{tN})-D_{c}(eA^{cNtK})e_{tN} \\
&=&-D_{c}(ee^{cK})-D_{c}(eA^{cNtK})e_{tN}\text{,}
\end{eqnarray*}%
\[
A^{cNtK}A_{cPtQ}=\delta _{P}^{N}e^{tK}e_{tQ}-\delta
_{Q}^{N}e^{tK}e_{tP}-\delta _{P}^{K}e^{tN}e_{tQ}+\delta _{Q}^{K}e^{tN}e_{tP}%
\text{,} 
\]%
\[
(\delta _{P}^{N}e^{tK}e_{tQ}-\delta _{Q}^{N}e^{tK}e_{tP})\overline{\Psi }%
(\gamma _{N}\frac{\sigma ^{PQ}}{2}+\frac{\sigma ^{PQ}}{2}\gamma _{N})\Psi =0%
\text{,} 
\]%
$e^{tM}e^{tN}\gamma _{M}\gamma _{N}=g^{tt}$ and (\ref{C-Cbar-Bracket}) lead
to%
\begin{eqnarray*}
\left\{ C(\overrightarrow{x}),\mathcal{D}_{t}^{\prime }(\overrightarrow{y}%
)\right\} &\simeq &-\frac{i}{2}(D_{c}(eA^{cNtK}) \\
&&+\frac{i}{2}e^{tP}\overline{\Psi }(\gamma _{P}\frac{\sigma ^{NK}}{2}+\frac{%
\sigma ^{NK}}{2}\gamma _{P})\Psi )e_{tN}\overline{\Psi }\gamma _{K}\delta (%
\overrightarrow{x}-\overrightarrow{y}) \\
&\simeq &-\frac{i}{2}M^{NK}e_{tN}\overline{\Psi }\gamma _{K}\delta (%
\overrightarrow{x}-\overrightarrow{y})\simeq 0\text{.}
\end{eqnarray*}

The same computation shows%
\[
\left\{ \overline{C}(\overrightarrow{x}),\mathcal{D}_{t}^{\prime }(%
\overrightarrow{y}\right\} \simeq -\frac{i}{2}M^{NK}\gamma _{K}\Psi
e_{tN}\delta (\overrightarrow{x}-\overrightarrow{y})\simeq 0. 
\]

Finally, a long and direct calculation shows that the Poisson bracket of the
smeared scalar constraint with itself is strongly equal to $(M\partial
_{a}M^{\prime }-M^{\prime }\partial _{a}M)$ times a linear combination of \
the primary constraints $C_{1}^{aKL}$, $C$ and $\overline{C}$ as%
\begin{equation}
\left\{ \mathcal{D}_{t}^{\prime }(M),\mathcal{D}_{t}^{\prime }(M^{\prime
})\right\} =\dint_{\Sigma }((M\partial _{a}M^{\prime }-M^{\prime }\partial
_{a}M)(F^{a}(C_{1})+G^{a}(C,\overline{C}))\simeq 0  \label{SC-SCPoisson}
\end{equation}%
where 
\begin{eqnarray*}
F^{a}(C_{1}) &=&-\frac{1}{2(d-2)g^{tt}}C_{1}^{aKL}A_{bKtL}e^{bQ}(me_{tQ}%
\overline{\Psi }\Psi \\
&&-e^{dI}e^{tP}\frac{i}{2}(\overline{\Psi }\gamma _{Q}\gamma _{P}\gamma
_{I}D_{d}\Psi -D_{d}\overline{\Psi }\gamma _{I}\gamma _{P}\gamma _{Q}\Psi ))
\\
&&+\frac{1}{2(d-2)(g^{tt})^{2}}C_{1}^{bKL}e_{bK}e^{aJ}e^{tP}e^{tM}(m%
\overline{\Psi }(\gamma _{L}\gamma _{P}\gamma _{J}\gamma _{M}+\gamma
_{M}\gamma _{J}\gamma _{P}\gamma _{L})\Psi \\
&&-e^{dI}i(\overline{\Psi }\gamma _{L}\gamma _{P}\gamma _{J}\gamma
_{M}\gamma _{I}D_{d}\Psi -D\overline{\Psi }\gamma _{I}\gamma _{M}\gamma
_{J}\gamma _{P}\gamma _{L}\Psi ))
\end{eqnarray*}%
and 
\begin{eqnarray*}
G^{a}(C,\overline{C}) &=&\frac{1}{(g^{tt})^{2}}e^{tN}e^{tM}e^{aK}(im(C\gamma
_{N}\gamma _{K}\gamma _{M}\Psi +\overline{\Psi }\gamma _{N}\gamma _{K}\gamma
_{M}\overline{C}) \\
&&+e^{dI}(C\gamma _{N}\gamma _{K}\gamma _{M}\gamma _{I}D_{d}\Psi -D_{d}%
\overline{\Psi }\gamma _{I}\gamma _{N}\gamma _{K}\gamma _{M}\overline{C}))%
\text{.}
\end{eqnarray*}

Note that as opposed in the pure gravity where the Poisson bracket of scalar
constraint with itself vanishes strongly \cite{Lagraa}, in presence of the
fermionic matter, this Poisson bracket vanishes only weakly. This shows that
the set of constraints is complete and closed meaning that the total
Hamiltonian (\ref{hAMILTON1}) is consistent.

\section{\protect\bigskip The Dirac brackets}

Before starting the constraint processing, let us note that instead of
directly analyzing the total Hamiltonian (\ref{TOTALHAM}) we can take the
non-dynamic part of the connection $\omega _{2aKL}$ and its conjugate moment 
$\mathcal{P}_{2}^{aKL}$ as primary constraints satisfying

\begin{equation}
\left\{ \omega _{2aKL}(\overrightarrow{x}),\mathcal{P}_{2}^{bPQ}(%
\overrightarrow{y})\right\} =P_{2KaL}^{\text{ \ \ \ \ \ \ \ }PbQ}\delta (%
\overrightarrow{x}-\overrightarrow{y})  \label{Primary2-bracket}
\end{equation}%
and contribute to the total Hamiltonian by adding the term%
\[
\int_{\Sigma }\left( \omega _{2aKL}\frac{\mathcal{B}_{2}^{aKL}}{2}+\mathcal{P%
}_{2}^{aKL}\frac{\mathcal{A}_{2aKL}}{2}\right) 
\]%
where $\mathcal{B}_{2}^{aKL}$ and $\mathcal{A}_{2aKL}$ are Lagrange
multiplier fields.

In order to satisfy the Jacobi identities we project the brackets acting on
the projected elements of the phase space $\omega _{2aKL}$ and $\mathcal{P}%
_{2}^{aKL}$ as 
\begin{eqnarray*}
\left\{ \pi ^{\mu N},\omega _{2aKL}\right\} &=&\left\{ \pi ^{\mu
N},P_{2KaL}^{\text{ \ \ \ \ \ \ \ }PbQ}\omega _{bPQ}\right\} _{2}=\left\{
\pi ^{\mu N},P_{2KaL}^{\text{ \ \ \ \ \ \ \ }PbQ}\right\} _{2}\omega _{bPQ}
\\
&=&P_{2KaL}^{\text{ \ \ \ \ \ \ \ }cRS}\left\{ \pi ^{\mu N},P_{2cRS}^{\text{
\ \ \ \ \ \ \ }dNM}\right\} P_{2dNM}^{\text{ \ \ \ \ \ \ \ }PbQ}\omega
_{bPQ}=0
\end{eqnarray*}%
due to $P(\delta P)P=0$ for any projector $P.$ The same computation gives%
\[
\text{ }\left\{ \pi ^{\mu N},\mathcal{P}_{2}^{aKL}\right\} =0 
\]%
which show that the Poisson brackets between these constraints and the
constraints $\pi ^{\mu N}$, $C_{1}^{aKL}$, $C$ and $\overline{C}$ vanish.
The consistency of the constraint $\omega _{2aKL}$ leads to $\mathcal{A}%
_{2aKL}=0$ and the consistency of $\mathcal{P}_{2}^{aKL}$ determines the
multiplier fields%
\begin{eqnarray*}
\left\{ \omega _{2aKL},\mathcal{H}_{T}\right\} &=&\mathcal{A}_{2aKL}=0 \\
\left\{ \mathcal{P}_{2}^{aKL},\mathcal{H}_{T}\right\} &=&-\frac{\mathcal{B}%
_{2}^{aKL}}{2}+\left\{ \mathcal{P}_{2}^{aKL},H_{0}\right\} =0
\end{eqnarray*}%
where $H_{0}$ is expressed in term $\omega _{aKL}=\omega _{1aKL}+\omega
_{2aKL}$. The first-class constraints of the previous chapter take the same
form plus terms proportional to the constraint $\omega _{2aKL}$ that can be
ignored. As a result, the constraints $\omega _{2aKL}$ and $\mathcal{P}%
_{2}^{aKL}$ are of second-class (\ref{Primary2-bracket}) and their Poisson
brackets with the other constraints vanish and so can just be discarded from
the theory leading to the total Hamiltonian (\ref{TOTALHAM}).

In this section we consider the second-class constraints $\pi ^{aN}$, $%
C_{1}^{aKL}$, $C$ and $\overline{C}$ as strong equalities by eliminating
them from the theory leading to the reduced Hamiltonian

\begin{equation}
\mathcal{H}_{T}^{r}=-\int_{\Sigma }(\mathcal{D}_{t}^{r}+M^{rKL}\frac{\omega
_{tKL}}{2})  \label{HamiltonianFinal}
\end{equation}%
where 
\[
M^{rKL}=(\pi ^{tK}e_{t}^{L}-\pi ^{tL}e_{t}^{K})+2D_{a}(eA^{aKtL})+iee^{tI}%
\overline{\Psi }(\gamma _{I}\frac{\sigma ^{KL}}{2}+\frac{\sigma ^{KL}}{2}%
\gamma _{I})\Psi 
\]%
are the reduced Lorentz constraints and 
\[
\mathcal{D}_{t}^{r}=-eA^{aKbL}\frac{\Omega _{1abKL}}{2}+\frac{i}{2}ee^{aK}(%
\overline{\Psi }\gamma _{K}D_{a}\Psi -D_{a}\overline{\Psi }\gamma _{K}\Psi
)-em\overline{\Psi }\Psi 
\]%
is the reduced scalare constraint. The diffeomorphism constraint reduces to%
\[
\mathcal{D}_{sp}^{r}(\overrightarrow{N})=\dint_{\Sigma }(\pi ^{tK}\mathcal{L}%
_{\overrightarrow{N}}(e_{tK})+eA^{aKtL}\mathcal{L}_{\overrightarrow{N}%
}(\omega _{1aKL})+\frac{i}{2}ee^{tK}(\overline{\Psi }\gamma _{K}\mathcal{L}_{%
\overrightarrow{N}}\Psi -\mathcal{L}_{\overrightarrow{N}}\overline{\Psi }%
\gamma _{K}\Psi ))\text{.} 
\]

In this case the algebra of the first-class constraints must be computed in
terms of projected Dirac brackets defined from the projected Poisson
brackets \cite{Lagraa} as

\[
\left\{ A,B\right\} _{D}=\left\{ A,B\right\} -\left\{ A,C_{i}\right\}
\left\{ C_{i},C_{j}\right\} ^{-1}\left\{ C_{j},B\right\} 
\]%
where $C_{i}=\left( \pi ^{aN},C_{1}^{aKL},C,\overline{C}\right) $ are the
set of the second-class constraints. The non zero elements of the inverse
super matrix $\left\{ C_{i},C_{j}\right\} ^{-1}$ are given by 
\[
\left\{ \pi ^{bN}(\overrightarrow{x}),C_{1}^{aKL}(\overrightarrow{y}%
)\right\} ^{-1}=e^{-1}B_{bNtKal}\delta (\overrightarrow{x}-\overrightarrow{y}%
)=-\left\{ C_{1}^{aKL}(\overrightarrow{x}),\pi ^{bN}(\overrightarrow{y}%
)\right\} ^{-1}\text{,} 
\]%
\[
\left\{ \pi ^{bN}(\overrightarrow{x}),C_{A}(\overrightarrow{y})\right\}
^{-1}=0=\left\{ \pi ^{bN}(\overrightarrow{x}),\overline{C}_{A}(%
\overrightarrow{y})\right\} ^{-1}\text{,} 
\]%
\[
\left\{ C_{A}(\overrightarrow{x}),\overline{C}_{B}(\overrightarrow{y}%
)\right\} _{+}^{-1}=i(eg^{tt})^{-1}e^{tI}\gamma _{IAB}\delta (%
\overrightarrow{x}-\overrightarrow{y})=\left\{ \overline{C}_{B}(%
\overrightarrow{y}),C_{A}(\overrightarrow{x})\right\} _{+}^{-1}\text{,} 
\]%
\begin{eqnarray*}
\left\{ C_{A}(\overrightarrow{x}),C_{1}^{aKL}(\overrightarrow{y})\right\}
^{-1} &=&-\frac{1}{2}(eg^{tt})^{-1}e^{tI}B_{bNtKal}A^{bNtM}(\gamma
_{I}\gamma _{M}\Psi )_{A}\delta (\overrightarrow{x}-\overrightarrow{y}) \\
&=&\left\{ C_{1}^{aKL}(\overrightarrow{y}),C_{A}(\overrightarrow{x})\right\}
^{-1}\text{,}
\end{eqnarray*}%
\begin{eqnarray*}
\left\{ \overline{C}_{A}(\overrightarrow{x}),C_{1}^{aKL}(\overrightarrow{y}%
)\right\} ^{-1} &=&-\frac{1}{2}(eg^{tt})^{-1}e^{tI}B_{bNtKal}A^{bNtM}(%
\overline{\Psi }\gamma _{M}\gamma _{I})_{A}\delta (\overrightarrow{x}-%
\overrightarrow{y}) \\
&=&\left\{ C_{1}^{aKL}(\overrightarrow{y}),\overline{C}_{A}(\overrightarrow{x%
})\right\} ^{-1}
\end{eqnarray*}%
and%
\begin{eqnarray*}
\left\{ C_{1}^{aKL}(\overrightarrow{x}),C_{1}^{bPQ}(\overrightarrow{y}%
)\right\} ^{-1} &=&-\frac{i}{4}%
(eg^{tt})^{-1}e^{tI}B_{dNtKal}A^{dNtM}B_{cJtPbQ}A^{cJtR}\times \\
&&\overline{\Psi }(\gamma _{M}\gamma _{I}\gamma _{R}-\gamma _{R}\gamma
_{I}\gamma _{M})\Psi \delta (\overrightarrow{x}-\overrightarrow{y})
\end{eqnarray*}%
from which we deduce the following non-zero Dirac brackets of the reduced
phase space $e_{aN}$, $\omega _{1aKL}$, $e_{tK\text{ }}$, $\pi ^{tK}$, $\Psi 
$and $\overline{\Psi }$ as%
\[
\left\{ e_{tN}(\overrightarrow{x}),\pi ^{tM}(\overrightarrow{y})\right\}
_{D}=\delta _{N}^{M}\delta (\overrightarrow{x}-\overrightarrow{y})\text{, }%
\left\{ e_{aN}(\overrightarrow{x}),\omega _{1bKL}(\overrightarrow{y}%
)\right\} _{D}=e^{-1}B_{aNtKbL}\delta (\overrightarrow{x}-\overrightarrow{y})%
\text{,} 
\]%
\begin{equation}
\left\{ \omega _{1aKL}(\overrightarrow{x}),\omega _{1bPQ}(\overrightarrow{y}%
)\right\} _{D}=\left\{ C_{1}^{aKL}(\overrightarrow{x}),C_{1}^{bPQ}(%
\overrightarrow{y})\right\} ^{-1}\text{,}  \label{DIRBRACon-Con}
\end{equation}%
\begin{equation}
\left\{ \Psi _{A}(\overrightarrow{x}),\omega _{1aKL}(\overrightarrow{y}%
)\right\} _{D}=\left\{ C_{A}(\overrightarrow{x}),C_{1}^{aKL}(\overrightarrow{%
y})\right\} ^{-1}=-\left\{ \omega _{1aKL}(\overrightarrow{y}),\Psi _{A}(%
\overrightarrow{x})\right\} _{D}\text{,}  \label{DIRBRACon-Psi}
\end{equation}

\begin{equation}
\left\{ \overline{\Psi }_{A}(\overrightarrow{x}),\omega _{1aKL}(%
\overrightarrow{y})\right\} _{D}=\left\{ \overline{C}_{A}(\overrightarrow{x}%
),C_{1}^{aKL}(\overrightarrow{y})\right\} ^{-1}=-\left\{ \omega _{1aKL}(%
\overrightarrow{y}),\overline{\Psi }_{A}(\overrightarrow{x})\right\} _{D}
\label{DIRBRACon-PsiB}
\end{equation}%
and 
\begin{equation}
\left\{ \Psi _{A}(\overrightarrow{x}),\overline{\Psi }_{B}(\overrightarrow{y}%
)\right\} _{+D}=-\left\{ C_{A}(\overrightarrow{x}),\overline{C}_{B}(%
\overrightarrow{y})\right\} _{+}^{-1}=\left\{ \overline{\Psi }_{B}(%
\overrightarrow{y}),\Psi _{A}(\overrightarrow{x})\right\} _{+D}\text{.}
\label{DIRBRAPsi-PsiB}
\end{equation}

A direct computation leads to the Dirac bracket between the spatial
diffeomorphism constraints as%
\begin{equation}
\left\{ \mathcal{D}_{sp}^{r}(\overrightarrow{N}),\mathcal{D}_{sp}^{r}(%
\overrightarrow{N^{\prime }})\right\} _{D}=\mathcal{D}_{sp}^{r}(\left[ 
\overrightarrow{N},\overrightarrow{N^{\prime }}\right] )
\label{RedDiff-RedDiff}
\end{equation}%
and to the spacial diffeomorphism transformations of the reduced phase space
elements as

\[
\delta e_{\mu N}=\left\{ e_{\mu N},\mathcal{D}_{sp}^{r}(\overrightarrow{N}%
)\right\} _{D}=\mathcal{L}_{\overrightarrow{N}}e_{\mu N}\text{, }\delta
\omega _{1aKL}=\left\{ \omega _{1aKL},\mathcal{D}_{sp}^{r}(\overrightarrow{N}%
)\right\} _{D}=\mathcal{L}_{1\overrightarrow{N}}\omega _{1aKL}\text{,} 
\]%
\[
\delta \Psi _{A}=\left\{ \Psi _{A},\mathcal{D}_{sp}^{r}(\overrightarrow{N}%
)\right\} _{D}=\mathcal{L}_{\overrightarrow{N}}\Psi _{A}\text{ and }\delta 
\overline{\Psi }_{A}=\left\{ \overline{\Psi }_{A},\mathcal{D}_{sp}^{r}(%
\overrightarrow{N})\right\} _{D}=\mathcal{L}_{\overrightarrow{N}}\overline{%
\Psi }_{A} 
\]%
from which we deduce that the scalar and Lorentz constraints transform like
scalar densities of weight one%
\begin{eqnarray}
\left\{ \mathcal{D}_{sp}^{r}(\overrightarrow{N}),\mathcal{D}_{t}^{r}\right\}
_{D} &=&-\mathcal{L}_{\overrightarrow{N}}(\mathcal{D}_{t}^{r})=-\partial
_{a}(N^{a}\mathcal{D}_{t}^{r})  \nonumber \\
&\Longrightarrow &\left\{ \mathcal{D}_{sp}^{r}(\overrightarrow{N}),\mathcal{D%
}_{t}^{r}(M)\right\} _{D}=\mathcal{D}_{t}^{r}(\mathcal{L}_{\overrightarrow{N}%
}(M))  \label{RedDiff-Redscal}
\end{eqnarray}%
and%
\begin{eqnarray}
\left\{ \mathcal{D}_{sp}^{r}(\overrightarrow{N}),M^{rKL}\right\} _{D} &=&-%
\mathcal{L}_{\overrightarrow{N}}M^{rKL}=-\partial _{a}(N^{a}M^{rKL}) 
\nonumber \\
&\Longrightarrow &\left\{ \mathcal{D}_{sp}^{\prime }(\overrightarrow{N}),%
\mathcal{M}^{r}(\theta )\right\} _{D}=\mathcal{M}^{r}(\mathcal{L}_{%
\overrightarrow{N}}(\theta ))\text{.}  \label{Reddiff-RedLor}
\end{eqnarray}

The reduced phase space elements transform under $\mathcal{M}^{r}(\theta )$
as infinitesimal gauge transformations

\[
\delta e_{\mu N}=\left\{ e_{\mu N},\mathcal{M}^{r}(\theta )\right\} _{D}=%
\mathcal{\theta }_{N}^{\text{ \ }M}e_{\mu M}\text{, }\delta \omega
_{1aKL}=\left\{ \omega _{1aKL},\mathcal{M}^{r}(\theta )\right\}
_{D}=-D_{1a}\theta _{KL}\text{,} 
\]%
\[
\delta \Psi =\left\{ \Psi ,\mathcal{M}^{r}(\theta )\right\} _{D}=\frac{%
\sigma ^{KL}}{2}\Psi \theta _{KL}\text{, }\delta \overline{\Psi }=\left\{ 
\overline{\Psi },\mathcal{M}^{r}(\theta )\right\} _{D}=-\overline{\Psi }%
\frac{\sigma ^{KL}}{2}\theta _{KL} 
\]%
which imply that $M^{rKL}$ transforms like a contravariant tensor\ 

\[
\left\{ M^{rKL},\mathcal{M}^{r}(\theta )\right\} _{D}=\theta _{\text{ \ }%
P}^{K}M^{rPL}+\theta _{\text{ \ }P}^{L}M^{rKP} 
\]%
leading to the $so(1,d-1)$ Lie algebra. From \ the transformations of the
reduced phase space the smeared scalar constraint $\mathcal{D}_{t}^{r}(M)$
transforms like a scalar%
\begin{equation}
\left\{ \mathcal{M}^{r}(\theta ),\mathcal{D}_{t}^{r}(M)\right\} _{D}=0\text{.%
}  \label{RedLor-Scal}
\end{equation}

The Dirac bracket between the smeared scalar constraints consist of the
following three parts: the first part concerns the purely gravitational part 
\begin{eqnarray*}
&&\left\{ -\dint_{\Sigma }MeA^{aKbL}\frac{\Omega _{1abKL}}{2},-\dint_{\Sigma
}M^{\prime }eA^{cNdM}\frac{\Omega _{1cdNM}}{2}\right\} _{D} \\
&=&-\dint_{\Sigma }\partial _{a}M\partial _{c}M^{\prime }\frac{i}{4g^{tt}}%
ee^{aK}e^{cN}e^{tI}\overline{\Psi }(\gamma _{K}\gamma _{I}\gamma _{N}-\gamma
_{N}\gamma _{I}\gamma _{K})\Psi \\
&&+\dint_{\Sigma }(M\partial _{a}M^{\prime }-M^{\prime }\partial _{a}M) \\
&&\frac{i}{4g^{tt}}e^{tI}e^{aK}B_{dRtNdM}A^{dRtS}D_{c}(eA^{cNdM})\overline{%
\Psi }(\gamma _{K}\gamma _{I}\gamma _{S}-\gamma _{S}\gamma _{I}\gamma
_{K})\Psi
\end{eqnarray*}%
which does not vanish as in the pure gravity. This is due to the non
vanishing Dirac brackets of the connection with itself in the presence of
the fermionic field. The fermionic part gives%
\begin{eqnarray*}
&&\left\{ \dint_{\Sigma }-M\frac{i}{2}ee^{aK}(\overline{\Psi }\gamma
_{K}D_{a}\Psi -D_{a}\overline{\Psi }\gamma _{K}\Psi ),\dint_{\Sigma
}-M^{\prime }\frac{i}{2}ee^{cN}(i\overline{\Psi }\gamma _{N}D_{c}\Psi -iD_{c}%
\overline{\Psi }\gamma _{N}\Psi )\right\} _{D} \\
&&+\left\{ \dint_{\Sigma }-M\frac{i}{2}ee^{aK}(\overline{\Psi }\gamma
_{K}D_{a}\Psi -D_{a}\overline{\Psi }\gamma _{K}\Psi ),\dint_{\Sigma
}M^{\prime }em\overline{\Psi }\Psi \right\} _{D}-(M\Leftrightarrow M^{\prime
}) \\
&&+\left\{ \dint_{\Sigma }Mem\overline{\Psi }\Psi ,\dint_{\Sigma }M^{\prime
}em\overline{\Psi }\Psi \right\} _{D} \\
&=&\dint_{\Sigma }(M\partial _{a}M^{\prime }-M^{\prime }\partial _{a}M)(%
\frac{i}{4g^{tt}}ee^{aK}e^{bL}e^{tI}(\overline{\Psi }\gamma _{K}\gamma
_{I}\gamma _{L}D_{b}\Psi -D_{b}\overline{\Psi }\gamma _{L}\gamma _{I}\gamma
_{K}\Psi ) \\
&&+\frac{i}{4g^{tt}}e^{aK}e^{tI}(\overline{\Psi }\gamma _{K}\gamma
_{I}\gamma _{L}D_{b}(ee^{bL}\Psi )-D_{b}(ee^{bL}\overline{\Psi })\gamma
_{L}\gamma _{I}\gamma _{K}\Psi )-\frac{em}{g^{tt}}e_{M}^{t}e^{aM}\overline{%
\Psi }\Psi \\
&&+\frac{1}{8g^{tt}}(e^{aK}e^{bL}e^{tI}\overline{\Psi }(\gamma _{L}\frac{%
\gamma ^{P}\gamma ^{Q}}{4}+\frac{\gamma ^{P}\gamma ^{Q}}{4}\gamma _{L})\Psi
B_{dRtPbQ}A^{dRtM}\overline{\Psi }(\gamma _{K}\gamma _{I}\gamma _{M}-\gamma
_{M}\gamma _{I}\gamma _{K})\Psi )) \\
&&-\dint_{\Sigma }\frac{i}{4g^{tt}}\partial _{a}M\partial _{c}M^{\prime
}ee^{aK}e^{cN}e^{tI}\overline{\Psi }(\gamma _{K}\gamma _{I}\gamma
_{N}-\gamma _{N}\gamma _{I}\gamma _{K})\Psi
\end{eqnarray*}%
and the third part gives%
\begin{eqnarray*}
&&\left\{ -\dint_{\Sigma }MeA^{aKbL}\frac{\Omega _{1abKL}}{2},\dint_{\Sigma
}M^{\prime }\frac{i}{2}ee^{cN}(\overline{\Psi }\gamma _{N}D_{c}\Psi -D_{c}%
\overline{\Psi }\gamma _{N}\Psi )-M^{\prime }e\overline{m\Psi }\Psi \right\}
_{D} \\
-(M &\Leftrightarrow &M^{\prime })=\dint_{\Sigma }\frac{i}{2g^{tt}}\partial
_{a}M\partial _{c}M^{\prime }ee^{aK}e^{cN}e^{tI}\overline{\Psi }(\gamma
_{K}\gamma _{I}\gamma _{N}-\gamma _{N}\gamma _{I}\gamma _{K})\Psi \\
&&+\dint_{\Sigma }(M\partial _{a}M^{\prime }-M^{\prime }\partial _{a}M)(-%
\frac{i}{4g^{tt}}ee^{aK}e^{bL}e^{tI}(\overline{\Psi }\gamma _{K}\gamma
_{I}\gamma _{L}D_{b}\Psi -D_{b}\overline{\Psi }\gamma _{L}\gamma _{I}\gamma
_{K}\Psi ) \\
&&-\frac{1}{8g^{tt}}ee^{aK}e^{bL}e^{tI}\overline{\Psi }(\gamma _{L}\frac{%
\sigma ^{PQ}}{2}+\frac{\sigma ^{PQ}}{2}\gamma _{L})\Psi B_{dRtPbQ}A^{dRtM}%
\overline{\Psi }(\gamma _{K}\gamma _{I}\gamma _{M}-\gamma _{M}\gamma
_{I}\gamma _{K})\Psi \\
&&-\frac{i}{4g^{tt}}e^{aK}e^{tI}(\overline{\Psi }\gamma _{K}\gamma
_{I}\gamma _{L}D_{b}(ee^{bL}\Psi )-D_{b}(ee^{bL}\overline{\Psi })\gamma
_{L}\gamma _{I}\gamma _{K}\Psi )+\frac{em}{g^{tt}}e_{M}^{t}e^{aM}\overline{%
\Psi }\Psi ) \\
&&-\dint_{\Sigma }(M\partial _{a}M^{\prime }-M^{\prime }\partial _{a}M)\times
\\
&&(\frac{i}{4g^{tt}}e^{tI}e^{aK}B_{eRtNdM}A^{eRtS}D_{c}(eA^{cNdM})\overline{%
\Psi }(\gamma _{K}\gamma _{I}\gamma _{S}-\gamma _{S}\gamma _{I}\gamma
_{K})\Psi )\text{.}
\end{eqnarray*}

These three parts cancel each other leading to%
\[
\left\{ \mathcal{D}_{t}^{r}(M),\mathcal{D}_{t}^{r}(M^{\prime })\right\}
_{D}=0 
\]%
which shows that the Dirac bracket of the reduced scalar constraints
strongly vanishes. This shows that the reduced first-class constraints
satisfy a closed algebra with structure constants.

It is easy to show that the physical degrees of freedom of the reduced phase
space match with those of the $d-$dimensional general relativity coupled
with the fermionic field.

Note that the reduced first-class constraints are polynomial but not the
Dirac brackets of the reduced phase space elements. In addition, the Dirac
brackets of the dynamic spacial connection with itself and with the
fermionic field do not vanish.

To get a new reduced phase space which is canonical with respect to the
Dirac brackets, we consider the following canonical transformation 
\begin{eqnarray}
e_{aN}\text{ } &\longrightarrow &\text{ }e_{aN}\text{,}  \nonumber \\
\omega _{1aKL}\text{ } &\longrightarrow &\mathcal{P}^{cN}(e,\omega _{1},\Psi
,\overline{\Psi })=eB^{cNtKal}\omega _{1aKL}+\frac{i}{2}eA^{cNtK}\overline{%
\Psi }\gamma _{K}\Psi \text{,}  \nonumber \\
\Psi \text{ } &\longrightarrow &\text{ }\Psi \text{ , }  \nonumber \\
\overline{\Psi }\text{ } &\longrightarrow &\Pi (e,\overline{\Psi })=iee^{tI}%
\overline{\Psi }\gamma _{I}  \label{Canonical-Trans}
\end{eqnarray}%
which results from the fact that the new variables $e_{aN}$, $\mathcal{P}%
^{cN}$,$\Psi $ and $\Pi $ of the reduced phase space result from the new
symplectic\ form of the action which differs from the one of (\ref%
{LagrangianF}) by a total derivative with respect of time%
\begin{eqnarray*}
&&\dint_{\mathcal{M}}\left( eA^{aKtL}\partial _{t}\omega _{1aKL}+ee^{tI}%
\frac{i}{2}(\overline{\Psi }\gamma _{I}\partial _{t}\Psi -\frac{i}{2}%
(\partial _{t}\overline{\Psi })\gamma _{I}\Psi )\right) \\
&=&\dint_{\mathcal{M}}\left( eB^{cNtKaL}\omega _{1aKL}+\frac{i}{2}eA^{cNtK}%
\overline{\Psi }\gamma _{K}\Psi )\partial _{t}e_{cN}+iee^{tI}\overline{\Psi }%
\gamma _{I}\partial _{t}\Psi \right) \text{.}
\end{eqnarray*}

Note that the the rank of the projector $P_{1KaL}^{\text{ \ \ \ \ \ \ \ }%
PdQ} $ is $d(d-1)$, implying that the number of the independent components
of $\omega _{1aKL}$ is equal to that of the co-tetrad $e_{aN}$ and that of $%
\mathcal{P}^{aN}$.

The primary constraints become 
\begin{eqnarray}
\pi ^{cN} &\longrightarrow &C^{aN}=\mathcal{P}^{cN}-eB^{cNtKal}\omega
_{1aKL}-\frac{i}{2}eA^{cNtK}\overline{\Psi }\gamma _{K}\Psi \simeq 0\text{,}
\nonumber \\
C_{1}^{aKL} &\longrightarrow &C_{1}^{aKL}=\mathcal{P}_{1}^{aKL}\simeq 0,%
\text{ }  \nonumber \\
C &\longrightarrow &C=\Pi -iee^{tI}\overline{\Psi }\gamma _{I}\simeq 0\text{
and}  \nonumber \\
\overline{C}\text{ } &\longrightarrow &\overline{C}=\overline{\Pi }\simeq 0
\label{Canonical-Ph-Sp}
\end{eqnarray}%
but the super matrix $\left\{ C_{i},C_{j}\right\} $ and its inverse $\left\{
C_{i},C_{j}\right\} ^{-1}$ keep the same form. This new reduced phase space
is canonical in the sense that their non-zero Dirac brackets are given by%
\begin{eqnarray}
\left\{ e_{aN}(\overrightarrow{x}),\mathcal{P}^{bM}(\overrightarrow{y}%
)\right\} _{D} &=&\delta _{a}^{b}\delta _{N}^{M}\delta (\overrightarrow{x}-%
\overrightarrow{y})\text{,}  \nonumber \\
\left\{ \Psi _{A}(\overrightarrow{x}),\Pi _{B}(\overrightarrow{y})\right\}
_{+D} &=&\delta _{AB}\delta (\overrightarrow{x}-\overrightarrow{y})\text{.}
\label{Canonical-Dir-Brac}
\end{eqnarray}

These canonical relations can be obtained either by using directly the Dirac
brackets with the constraints (\ref{Canonical-Ph-Sp}) or by using the Dirac
brackets between the previous reduced phase space variables.

The inverse of the canonical transformation (\ref{Canonical-Trans}) gives $%
\omega _{1aKL}$ and $\overline{\Psi }$ as functions of the canonical phase
space 
\begin{equation}
\omega _{1aKL}(e,\mathcal{P},\Psi ,\Pi )=\mathcal{P}%
^{bM}e^{-1}B_{bMtKaL}-B_{bMtKaL}\frac{i}{2}A^{bMtN}\overline{\Psi }\gamma
_{N}\Psi  \label{NewSPconnectio}
\end{equation}%
and 
\begin{equation}
\overline{\Psi }(e,\Pi )=-i(eg^{tt})^{-1}e^{tI}\Pi \gamma _{I}
\label{NewFerbar}
\end{equation}%
leading to polynomial constraints of first-class of the form%
\[
\mathcal{D}_{sp}^{r}(\overrightarrow{N})=\dint_{\Sigma }\left( \pi ^{tK}%
\mathcal{L}_{\overrightarrow{N}}(e_{tK})+\mathcal{P}^{aK}\mathcal{L}_{%
\overrightarrow{N}}(e_{aK})+\Pi \mathcal{L}_{\overrightarrow{N}}(\Psi
)\right) 
\]%
and 
\[
\mathcal{M}^{r}(\theta )=\dint_{\Sigma }\left( \partial _{a}(eA^{aKtL})+%
\frac{1}{2}(\mathcal{P}^{aK}e_{a}^{L}-\mathcal{P}^{aL}e_{a}^{K})+\Pi \frac{%
\sigma ^{KL}}{2}\Psi \right) \theta _{KL}=\dint_{\Sigma }M^{rKL}\frac{\theta
_{KL}}{2} 
\]%
but the scalar constraint $\mathcal{D}_{t}^{r}(M)$ take a non polynomial
form.

A calculation analogous to that performed in \cite{Lagraa}, to demonstrate
Jacobi's identities, shows that the dynamic spatial connection (\ref%
{NewSPconnectio}) satisfies the same projected Dirac brackets as that of the
reduced phase (\ref{DIRBRACon-Con}-\ref{DIRBRACon-PsiB}). This leads to the
same closed algebra of reduced first-class constraints expressed in terms of
the new reduced phase space equipped with the canonical Dirac brackets (\ref%
{Canonical-Dir-Brac}).

\section{Conclusion}

We have showed in this paper that a modified action of the tetrad-gravity
coupled with the fermionic field where the non-dynamic part of the
connection is fixed to zero leads to a consistent Hamiltonian formalism in
any dimension $d\geq 3$ free of the Barbero-Immirzi parameter. Contrarily to
the different Hamiltonian formalisms based on the ADM construction where the
algebra of the first-class constraint is closed with structure functions,
here we are in presence of an algebra of first-class constraints which
closes with structure constants. Therefore, this algebra generates true
Lie-group transformations expressing the local invariance of the
tetrad-gravity coupled with the fermionic matter under Lorentz
transformations $\mathcal{M}^{r}(\theta )$\ and diffeomorphism whose smeared
generators are the scalar constraint $\mathcal{D}_{t}(M)$ and the vector
constraint $\mathcal{D}_{sp}(\overrightarrow{N})$. The scalar function $M$
and the spatial vector field $\overrightarrow{N}$ may be interpreted as the
usual lapse and shift respectively although they do not result from the
A.D.M. formalism \cite{Arnowitt}. The absence of the structure functions is
due to the fact that the scalar function $M$ and the spatial vector field $%
\overrightarrow{N}$ are introduced here only as smeared functions
independently to the decomposition A.D.M of the tangent space which requires
the metric which appears in the structure functions \cite{Hojman}.

Although the Dirac brackets of the reduced phase space between the dynamic
spatial connection with itself and with the fermionic fields are
complicated, the first-class constraints satisfy a closed algebra with
structure constant analogous to the one of the pure gravity. We have also
showed that a canonical transformation leads to a new reduced phase space
endowed with Dirac brackets having a canonical form leading to the same
algebra with structure constants.

Now we can investigate the contribution of the torsion to the connections.
For that, we solve the Hamiltonian equations in terms of the Dirac brackets.
The temporal evolution of the tetrad components is given by%
\begin{eqnarray*}
\frac{de_{aN}}{dt} &=&\left\{ e_{aN},\mathcal{H}_{T}^{r}\right\}
_{D}=D_{a}e_{tN}-B_{aNtK\mu L}e^{\mu I}\frac{ie}{2}\overline{\Psi }(\gamma
_{I}\frac{\sigma ^{KL}}{2}+\frac{\sigma ^{KL}}{2}\gamma _{I})\Psi \\
&&+B_{aNtKtL}M^{rKL}-\omega _{tN}^{\text{ \ \ \ \ }M}e_{aM}
\end{eqnarray*}%
where the constraint $M^{rKL}$ is expressed with $\pi ^{tN}=0$. Modulo the
constraint $M^{rKL}$, which is the equation obtained by varying the action
by connection $\omega _{tKL}$, we get%
\begin{equation}
D_{t}e_{aN}-D_{a}e_{tN}=-B_{aNtK\mu L}e^{\mu I}\frac{ie}{2}\overline{\Psi }%
(\gamma _{I}\frac{\sigma ^{KL}}{2}+\frac{\sigma ^{KL}}{2}\gamma _{I})\Psi 
\text{.}  \label{Equation1tortion}
\end{equation}

The solutions of this equation and $M^{rKL}=0$ give the expression of the
connection in terms of the tetrad and the fermionic field. First, the
equation 
\begin{equation}
M^{rKL}=2D_{a}(eA^{aKtL})+iee^{tI}\overline{\Psi }(\gamma _{I}\frac{\sigma
^{KL}}{2}+\frac{\sigma ^{KL}}{2}\gamma _{I})\Psi =0  \label{Equation2tortion}
\end{equation}%
can be solved for the torsion by splitting the connection as $\omega _{\mu
KL}=\widetilde{\omega }_{\mu KL}+C_{\mu KL}$ \cite{Pedan} where $\widetilde{%
\omega }_{\mu KL}=e_{\upsilon K}\nabla _{\mu }e_{L}^{\upsilon }=e_{\upsilon
K}(\partial _{\mu }e_{L}^{\upsilon }+\Gamma _{\mu \alpha }^{\nu
}e_{L}^{\alpha })$ is the free-torsion part of the connnection compatible
with the tetrad and $C_{\nu KL}$ is the contorsion. Since $P_{2KaL}^{\text{
\ \ \ \ \ \ \ }NbM}\omega _{bNM}^{s}$ is the solution of the homogeneous
equation (\ref{Equation2tortion}), the general solution is $\omega
_{aKL}^{s}=\widetilde{\omega }_{aKL}+C_{aKL}=P_{1KaL}^{\text{ \ \ \ \ \ \ \ }%
NbM}\omega _{bNM}^{s}+P_{2KaL}^{\text{ \ \ \ \ \ \ \ }NbM}\omega _{bNM}^{s}$
and so the equations (\ref{Equation1tortion}) and (\ref{Equation2tortion})
are expressed with the full solution $\omega _{aKL}^{s}$. The solution of (%
\ref{Equation2tortion}) is given by%
\begin{eqnarray}
D_{a}e_{bN}-D_{b}e_{aN} &=&C_{aN}^{\text{ \ \ \ \ }M}e_{bM}-C_{bN}^{\text{ \
\ \ \ }M}e_{aM}=  \nonumber \\
&&B_{aNbKtL}\frac{ie}{2}e^{tI}\overline{\Psi }(\gamma _{I}\frac{\sigma ^{KL}%
}{2}+\frac{\sigma ^{KL}}{2}\gamma _{I})\Psi  \nonumber \\
&&+B_{aNbKcL}\frac{ie}{2}e^{cI}\overline{\Psi }(\gamma _{I}\frac{\sigma ^{KL}%
}{2}+\frac{\sigma ^{KL}}{2}\gamma _{I})\Psi  \label{Solution2tortion}
\end{eqnarray}%
where, as a consequence of (A.8), the second term of the right hand side is
the solution of the homogeneous equation (\ref{Equation2tortion}). (\ref%
{Equation1tortion}) and (\ref{Solution2tortion}) are exactly the solutions
of the equations obtained by varying with respect of the connection the
action of gravity coupled minimally with the fermions.

We end this work by noting that the main purpose of the canonical formalism
of the gravity is its canonical quantization which is made off-shell with
the connection not with its free-torsion part and the contortion which
result from the solutions of equation of motion.

\section{Appendix A}

\bigskip In this appendix, we collect the properties of the functions $A$, $%
B $ \cite{Lagraa}, \cite{Frolov} and the projectors used in the Hamiltonian
analysis.

\begin{eqnarray}
eA^{\mu K\nu L} &=&\frac{\delta }{\delta e_{\mu K}}ee^{\nu L}=\frac{1}{(d-2)!%
}\epsilon ^{I_{0}...I_{d-3}KL}e_{\mu _{0}I_{0}}...e_{\mu
_{d-3}I_{d-3}}\epsilon ^{\mu _{0}...\mu _{d-3}\mu \nu }  \nonumber \\
&=&e(e^{\mu K}e^{\nu L}-e^{\nu K}e^{\mu L})=-eA^{\nu K\mu L}=-eA^{\mu L\nu K}
\TCItag{A.1}
\end{eqnarray}

where $\epsilon _{I_{0...}I_{d-1}}$ are the components of the totally
antisymmetric Levi-Cevita symbol, $\epsilon _{0...d-1}=-\epsilon
^{0...d-1}=1 $.%
\begin{eqnarray}
eB^{\beta N\mu K\nu L} &=&\frac{1}{(d-3)!}\epsilon
^{I_{0}...I_{d-4}NKL}e_{\mu _{0}I_{0}}...e_{\mu _{d-4}I_{d-4}}\epsilon ^{\mu
_{0}...\mu _{d-4}\beta \mu \nu }  \nonumber \\
&=&e(e^{\beta N}A^{\mu K\nu L}+e^{\beta K}A^{\mu L\nu N}+e^{\beta L}A^{\mu
N\nu K})  \nonumber \\
&=&e(e^{\beta N}A^{\mu K\nu L}+e^{\mu N}A^{\nu K\beta L}+e^{\nu N}A^{\beta
K\mu L})=\frac{\delta }{\delta e_{\beta N}}eA^{\mu K\nu L}  \TCItag{A.2}
\end{eqnarray}%
with its inverse%
\begin{eqnarray}
B_{\mu N\nu K\alpha L} &=&\frac{1}{2}\left( e_{\mu N}\frac{A_{\nu K\alpha L}%
}{d-2}+e_{\nu N}\frac{A_{\alpha K\mu L}}{d-2}+e_{\alpha N}A_{\mu K\nu
L}\right)  \nonumber \\
&=&\frac{1}{2}\left( \frac{A_{\mu N\nu K}}{d-2}e_{\alpha L}+\frac{A_{\mu
L\nu N}}{d-2}e_{\alpha K}+A_{\mu K\nu L}e_{\alpha N}\right)  \TCItag{A.3}
\end{eqnarray}%
in the sense that 
\begin{equation}
B_{\mu N\nu K\alpha L}B^{\mu N\nu P\beta Q}=\delta _{\alpha }^{\beta
}(\delta _{K}^{P}\delta _{L}^{Q}-\delta _{L}^{P}\delta _{K}^{Q})  \tag{A.4}
\end{equation}%
and%
\begin{equation}
B_{\mu N\nu K\alpha L}B^{\rho M\sigma K\alpha L}=\delta _{N}^{M}(\delta
_{\mu }^{\rho }\delta _{\nu }^{\sigma }-\delta _{\nu }^{\rho }\delta _{\mu
}^{\sigma }).  \tag{A.5}
\end{equation}

We see from (A.3) that, unlike $A^{\mu K\nu L}$ and $B^{\mu N\nu K\alpha L}$
which are antisymmetric in interchange of two indices of the same nature, $%
B_{\mu N\nu K\alpha L}$ is only antisymmetric on the indices $\mu $ and $\nu 
$ and on the indices $K$and $L$. As a consequence of the antisymmetric of
the indices $\mu ,$ $\nu $ and $\alpha $ of $B^{\mu M\nu K\alpha L}$, (A.5)
gives 
\begin{equation}
B_{cNtKaL}B^{bMtKaL}=\delta _{N}^{M}\delta _{c}^{b}  \tag{A.6}
\end{equation}%
for $\sigma =\nu =t$ and

\begin{equation}
B_{cNdKaL}B^{bMtKaL}=0  \tag{A.7}
\end{equation}%
for $\sigma =t$ and $\nu =d$. (A.4) leads to%
\begin{equation}
B^{aNbPtQ}B_{aNbK\mu L}=\delta _{\mu }^{t}(\delta _{K}^{P}\delta
_{L}^{Q}-\delta _{L}^{P}\delta _{K}^{Q})  \tag{A.8}
\end{equation}

A straightforward computation shows that the contraction of $B_{cNdKaL}$
which one of the components $e^{cN}$, $e^{dN}$, $e^{aL}$, $e^{aK}$, $e^{tN}$%
, $e^{tK}$ or $e^{tL}$ vanishes. From (A.4), (A.5) and (A.6) we deduce that

\begin{equation}
P_{1KaL}^{\text{ \ \ \ \ \ \ \ }PdQ}=B_{bNtK\alpha L}B^{bNtPdQ}\text{ and }%
P_{2KaL}^{\text{ \ \ \ \ \ \ \ }PdQ}=\frac{1}{2}B_{bNcK\alpha L}B^{bNcPdQ} 
\tag{A.9}
\end{equation}%
are projectors:

\[
P_{1KaL}^{\text{ \ \ \ \ \ \ \ }PdQ}+P_{2KaL}^{\text{ \ \ \ \ \ \ \ }PdQ}=%
\frac{1}{2}\delta _{a}^{d}\left( \delta _{K}^{P}\delta _{L}^{Q}-\delta
_{L}^{P}\delta _{K}^{Q}\right) \text{,} 
\]

\[
P_{1KaL}^{\text{ \ \ \ \ \ \ \ }NbM}P_{1NbM}^{\text{ \ \ \ \ \ \ \ }%
PdQ}=P_{1KaL}^{\text{ \ \ \ \ \ \ \ }PdQ}\text{, }P_{2KaL}^{\text{ \ \ \ \ \
\ \ }NbM}P_{2NbM}^{\text{ \ \ \ \ \ \ \ }PdQ}=P_{2Kal}^{\text{ \ \ \ \ \ \ \ 
}PdQ} 
\]%
and%
\[
P_{1KaL}^{\text{ \ \ \ \ \ \ \ }NbM}P_{2NbM}^{\text{ \ \ \ \ \ \ \ }PdQ}=0. 
\]

Their ranks are given by their trace 
\[
\frac{1}{2}\delta _{d}^{a}\left( \delta _{P}^{K}\delta _{Q}^{L}-\delta
_{P}^{K}\delta _{P}^{L}\right) P_{1KaL}^{\text{ \ \ \ \ \ \ \ }PdQ}=d(d-1) 
\]%
and 
\[
\frac{1}{2}\delta _{d}^{a}\left( \delta _{P}^{K}\delta _{Q}^{L}-\delta
_{P}^{K}\delta _{P}^{L}\right) P_{2KaL}^{\text{ \ \ \ \ \ \ \ }PdQ}=\frac{1}{%
2}d(d-1)(d-3)\text{.} 
\]

Inserting $\omega _{2aKL}$ in the symplectic part of the action, we get 
\begin{eqnarray}
eA^{aKtL}\partial _{t}\omega _{2aKL} &=&\partial _{t}(eA^{aKtL}\omega
_{2aKL})-\partial _{t}(eA^{aKtL})\omega _{2aKL}  \nonumber \\
&=&eB^{bNtKaL}(\partial _{t}e_{bN})\omega _{2aKL}=0  \TCItag{A.10}
\end{eqnarray}%
which shows that $\omega _{2aKL}$ is non-dynamic part in the sense that it
does not contribute to the symplectic part of the action. Similarly for $%
D_{2a}\omega _{tKL}=P_{2KaL}^{\text{ \ \ \ \ \ \ }PdQ}D_{d}\omega _{tPQ}$

\[
eA^{aKtL}D_{2a}\omega _{tKL}=0\text{.} 
\]

\end{document}